\documentclass[12pt]{JHEP3}
\pdfoutput=1
\usepackage{graphicx}

\usepackage{epsfig}
\usepackage{amssymb}
\usepackage{bm}
\usepackage{amsmath}
\makeatletter
\@addtoreset{equation}{section}
\makeatother

\title{
Matrix Theory for Baryons:
An Overview of Holographic QCD for Nuclear Physics
}
\author{
Sinya Aoki$^{1}$, Koji Hashimoto$^{2}$ and  Norihiro Iizuka$^{3}$ \\

$^1$ {\it Graduate School of Pure and Applied Sciences,
University of Tsukuba, Ibaraki 305-8571, Japan}, and\\
{\it Center for Computational Sciences,
University of Tsukuba, Ibaraki 305-8577, Japan} \\
E-mail: \email{saoki(at)het.ph.tsukuba.ac.jp}\\

$^2$
{\it Mathematical Physics Lab., RIKEN Nishina Center,
Saitama 351-0198, Japan}\\
E-mail: \email{koji(at)riken.jp}\\ 

$^3$
{\it Theory Division, CERN, CH-1211 Geneva 23, Switzerland}\\
E-mail: \email{norihiro.iizuka(at)cern.ch}\\

}

\abstract{
We provide, for non-experts, a brief overview of holographic QCD and a review
of a recent proposal of matrix-description \cite{Hashimoto:2010je}
of multi-baryon systems in holographic QCD.
Based on the matrix model, we derive 
the baryon interaction at short distances in multi-flavor holographic QCD.  
We show that there is a very universal repulsive core of inter-baryon forces 
for generic number of flavors. This is consistent with a recent lattice
QCD analysis for $N_f=2, 3$ where repulsive core looks universal. 
We also provide a comparison of our results with the lattice QCD and the operator product expansion (OPE) analysis.

}

\preprint{
{\normalsize CERN-PH-TH-2012-072} \\
{\normalsize RIKEN-MP-42}\\
{\normalsize UTHEP-640}
}

\begin{document}
\setcounter{page}{1}


\noindent
\section{M-theory for Nuclear Physics?} 
\label{sec1}

What is ``M-theory" for nuclear physics?  Although the ``M-theory"\footnote{M is for mystery, mother, matrix. \cite{M-theory}} stands for a theory of
everything which unifies all string theories \cite{M-theory}, one can generalize the use of the word
``M-theory" not only for string theories but also for other subjects 
in physics. 
What is M-theory for nuclear physics, if exists?

This kind of question brings us to a bigger picture of relations between various subjects within physics,
so it is not of no use.  
The question, however, sounds ridiculous, because the answer for it is
obvious: The M-theory for nuclear physics is QCD, or more precisely, 
the Standard Model of elementary particles. Nucleons, which are the building blocks
of nuclei, are bound states of quarks and gluons in QCD. Supposing that  one could solve QCD
completely, in principle one should be able to derive all the properties of nuclei, which is nothing but 
the nuclear physics.
Therefore, in this sense, QCD is the M-theory for the nuclear physics. However, QCD is 
notorious as being difficult to solve, due to 
its strong coupling nature: the strong force makes quarks bound to each other.
Therefore we need a new tool for solving QCD to ``derive" nuclear physics. 
Once the new tool is available, we may then say that we ``understand" the 
real-world nuclear physics phenomena from M-theory.

\begin{figure}
\centering
\includegraphics[scale=0.4]{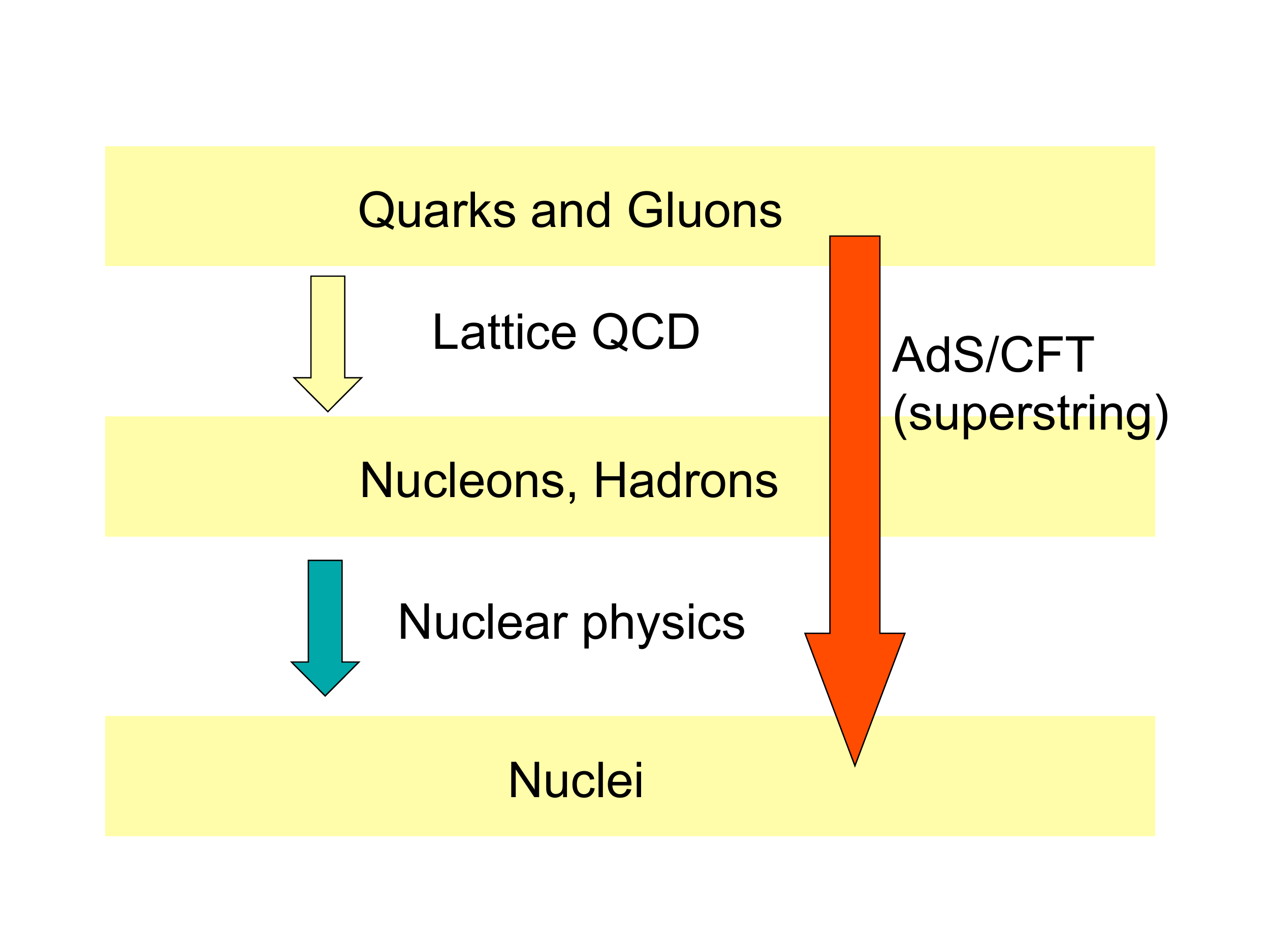}
\caption{A conceptual merit of the AdS/CFT correspondence, for a possible bridge between 
elementary particle physics, hadron physics and nuclear physics.}
\label{fig1}
\end{figure}

Since this new tool has been missing for long years in research, apparently we have a hierarchical
structure between studying perturbative QCD, nuclear physics and hadron physics (see Fig.~\ref{fig1}).
Standard nuclear physics starts with a quantum mechanics
of multi nucleons, with inter-nucleon potential (nuclear force) 
given by experiments, or by hand to match phenomena. The quantum mechanics Lagrangian becomes
\begin{eqnarray}
S = \int \! dt 
\left[
\sum_{s=1}^A \frac{M}{2} \left(\partial_t x^M_{(s)}(t)\right)^2
-\sum_{s_1\neq s_2}V[x_{(s_1)}^M-x_{(s_2)}^M] + \cdots
\right],
\label{nucac}
\end{eqnarray}
where we have $A$ nucleons whose locations are given by $x^M_{(s)}(t)$ with $s=1,\cdots,A$.
The first term is the kinetic term of the nucleons with mass $M$, 
while the second term is the nuclear force.
The problem lying in the unification of our concern is the fact that in nuclear physics the
nuclear force $V$ is given by experiments, and not by fundamental theory, {\it i.e.}, QCD.
In principle, the potential should have been
derived from QCD, as we all know that nucleons and hadrons are made of quarks and gluons --- 
but it is very difficult.

It is very recent that the nuclear force was calculated from QCD with use of numerical methods: 
lattice QCD \cite{Ishii:2006ec, Aoki:2009ji}.
The lattice QCD has accomplished a great success in hadron physics. In particular for hadron spectroscopy
and hadron interactions, the lattice QCD is now very close to the physical parameters of QCD, the real world.
Furthermore, there is a progress in this direction toward nuclear physics itself \cite{Yamazaki:2009ua}. 
Once the lattice QCD comes to deal with a system of multi-baryons, a part of nuclear physics becomes
accessible directly from QCD. 
A huge number of quark contractions in large nuclei, which requires almost unrealistically high  power of supercomputers, however, is a big obstacle in this direction. 
Furthermore, it is of course more ideal if we can understand physics without relying on the computers.  Unfortunately we have not yet reached that stage. 
Therefore, we are facing at a situation where the hadron physics and the nuclear physics are disconnected each other in a sense, due to a difficulty in solving the strongly coupled QCD.

At this occasion, the new tool using string theory comes into play. The renowned AdS/CFT 
correspondence \cite{Maldacena:1997re, Gubser:1998bc}
makes it possible to solve a certain limit of QCD-like gauge theories, and it offers a certain direct
path from QCD to nuclear physics. If one can derive an action like \eqref{nucac} from QCD, 
it can be regarded as an effective theory for nuclear physics derived from M-theory. 

In this paper, we review the recent progress along this direction 
as an application of the AdS/CFT correspondence. In \cite{Hashimoto:2010je}, 
two of the present authors
(K.H.~and N.I.), together with Piljin Yi, derived an action of a multi-baryon system, by using 
the AdS/CFT correspondence applied to large $N_c$ QCD. The action indeed has the form of 
\eqref{nucac}, and it serves as a candidate for the bridge between QCD and nuclear physics.

As it was derived from the large $N_c$ QCD, the action is written only with
two free parameters: the QCD scale and the QCD coupling. Therefore we can make a check of the
derived theory by just calculating various observables in nuclear physics with this action and compare those with experiments, to test the validity of the action, up to the approximations of 
the large $N_c$ and the strong coupling expansion.
Explicitly demonstrated in the literature are: 

(i) Baryon spectroscopy \cite{Hashimoto:2010je} 

(ii) Universal repulsive core of nucleons \cite{Hashimoto:2010je}

(iii) Three-body nuclear forces \cite{Hashimoto:2010ue} 

(iv) Spin statistics of baryons \cite{arXiv:1006.3612}

(v) Formation of atomic nuclei \cite{arXiv:1103.5688}

\vspace{5mm}

In all of these calculations, the results are qualitatively reasonable compared to the experiments\footnote{Within the same framework, 
using the flavor brane action, 
 it has been reported that a wider class of results are compatible with experiments: 
 baryon spectrum was originally derived in \cite{Hata:2007mb}, and charge radii of baryons \cite{Hashimoto:2008zw}, 
 suppression of multi-nucleon forces \cite{Hashimoto:2009as}, baryon spectra with three flavors \cite{Hashimoto:2009st}, {\it etc}.
 The repulsive core has been calculated in the same manner \cite{Hashimoto:2009ys}.
 See also some alternative approach given in \cite{Hong:2007kx}.}. As we will
 describe in this paper, there are a lot more physical observables which can be calculated
 in the framework.

The first aim of this paper is to give a review of the effective action for the multi-baryon system 
\cite{Hashimoto:2010je}, for non-experts of holographic methods. The second aim is to show a
new result on the short-distance force between baryons with multi flavors where the number of flavors $N_f$ is $N_f > 2$.

This paper is organized as follows. The first part of this paper is mainly a review. 
In section 2, we give a brief review of the status of holographic QCD, explaining the difference between the holographic QCD and real QCD, to emphasize what are 
remaining problems in holographic QCD. Then in section 3,
we shall explain the nuclear physics action derived in AdS/CFT correspondence, with emphasis on
its properties, new insights and connection to nuclear physics, for non-experts.  

The second part of this paper consists of new results. In section 4, 
we calculate the short distance inter-baryon forces
for the case of multi-flavors (the number of flavors larger than 2). 
We shall see that the repulsive core remains even for generic number of flavors,
thus find a universal repulsive core.
The result is consistent 
with recent lattice results with $N_f = 3$ where in most of the channels
there appears an inter-baryon repulsive potential. 
Therefore our result would also serve as another nontrivial consistency check.
This part of the paper includes technical details. Readers who know holographic QCD and the matrix model
approach of \cite{Hashimoto:2010je} can start with section 4 as it is written independently of section 2 and 3.
In the last section 5, we provide a review of the recent lattice results for multi-flavors and also the operator product expansion (OPE), 
as a comparison to
our holographic results.


\noindent
\section{Universal Problems in Holographic QCD}
\label{sec2}

For readers who are not familiar with the subject of the holographic QCD 
(the AdS/CFT correspondence applied to QCD),\footnote{In this article, and in most of articles in this field,
the word ``AdS/CFT" is equivalent to the word ``gauge/gravity'' or ``holography"  in use. 
We say ``bulk'' for gravity or string side calculation and ``boundary'' for the gauge theory side calculations.}
here in this section, we summarize important problems which are to be addressed 
in the holographic QCD. In particular, we make a stress on what are assumptions and what are
ignored in holographic QCD. This would make clear an importance and a validity of the AdS/CFT matrix
model approach to multi-baryon system and nuclear physics, which we shall review in the next section.

\subsection{How holographic QCD is different from QCD}

The holographic QCD is different from the real QCD. The holographic QCD, however, is very important setting 
as it provides us with a non-perturbative and analytic method to systematically study the strong coupling nature of gauge theories including QCD-like gauge theories. 
Therefore we first need to know why it is difficult to
directly apply the holography method to the real 
QCD.\footnote{Here keep in mind that we are talking about
top-down approach of holography from string theory. 
Any bottom-up approach, to write down higher-dimensional gravity
models as phenomenological models for QCD, does not have clear 
understanding on which gauge theory is dual to those bottom-up gravity models.  
Therefore, to be precise, we discuss top-down model only in the framework derived in string theory in this paper.
Bottom-up models are criticized only through their comparison to real QCD data, while
top-down models can be more concrete in criticisms as they are directly related to QCD through
string theory or D-brane construction.}

\subsubsection{Forced large $N_c$ and $\lambda$ limits}

The AdS/CFT correspondence in string theory is a conjecture
on the equivalence between a non-gravitational gauge theory and a {\it string theory in asymptotically AdS background}. 
Note that the duality is {\it not} between gauge theory and gravity, but rather string theory.  
Only when large $N_c$ and large $\lambda$ (which is a 't Hooft coupling of the gauge theory)
are taken, the sting theory side can be approximated by a gravity theory with background geometries of 
weakly curved spacetime. Low energy excitations of the string, such as gravitons, 
are light, while the long string itself becomes very heavy. This 
is almost an unique universal situation where one can describe low energy physics concretely 
in string theory as gravitational theory. 
Due to the technical difficulty of solving string theory in generically curved background, 
in many situation where we apply the holography to 
theories like QCD, 
the two limits, large $N_c$ and large $\lambda$, 
are forced so that we can approximate string theory as a gravity theory. 

It is this approximation which makes us face several difficulty 
in the comparison between the holographic QCD and realistic QCD or nuclear phenomena. 
According to old string models, 
hadrons with higher spins are stringy excitations. Holographic QCD follows and generalizes the
old string models based on QCD strings. 
In the AdS/CFT correspondence, the tension of the strings in the gravity side is ${\cal O}(\lambda)$,
so the stringy excitations become extremely heavy, and resultantly, 
the higher spin modes are parametrically heavy and decouple from 
the gravity excitations. 
This is the reason why in large $\lambda$, string theory is approximated as a gravitational theory. 
If we take $\lambda$ to infinity, however,   
the highest spin excitation of the system be graviton, which has spin 2, and all the stringy modes whose 
spin are bigger than 2 be infinitely heavy.  
On the other hand, in nuclear physics, there are 
many hadronic excitations whose spin are bigger than 2 and all of these higher spin hadronic excitations have the 
same order mass scale compared with lower spin excitations. 
Therefore, in holographic QCD within the gravitational approximation, 
we should keep in our mind that there could be a contradiction for a comparison with data caused by the missing degrees of freedom whose spin are bigger than 2. 

One may then wonder why we do not directly try to solve string theory in asymptotically AdS background, instead of using the 
gravity approximation. 
A problem is that a precise treatment of the fundamental strings in the curved geometry, {\it i.e.} the quantization
of the string, is
still missing in any formulations of string theory. We miss a fundamental tool to analyze the string side. 
This waits for a further development of methods to quantize strings in curved geometries.\footnote{
One major progress along this line would be a correspondence between vector models 
and higher-spin gauge 
theories \cite{Fradkin:1987ks}, as an explicit  toy model of the AdS/CFT correspondence in the $\lambda\rightarrow 0$ limit.} 

In addition, we have $N_c=3$ in realistic QCD,  while  any quantities are computed at the leading order of the $1/N_c$ expansion around $N_c = \infty$ in holographic QCD. So, in comparison to experiments, we expect, at least,   
33\% or more errors generically due to the large $N_c$ approximation. 
At present, computations of the sub-leading $1/N_c$ corrections, which correspond to that of
string loop corrections in the gravity side, are technically difficult. These all imply that the holography methods
are better applied to reveal some robust features of QCD, which are independent on the values of $N_c$, 
not to make a comparison to precision measurements.

We might wonder under what circumstance physical quantities could be more insensitive to $N_c$. 
In the confining phase of QCD, as the color degrees of freedom 
are confined, we cannot directly observe the number of colors.  We therefore naively might expect that physics might be independent on the values of $N_c$. On the other hand, 
physics in the deconfining phase would suffer more defects from the large $N_c$ limit.
Interestingly, however, there are many successful examples of the calculations in the deconfining phase  
for the shear viscosity \cite{viscosity}, quark energy loss \cite{energylosspaper} etc in the quark gluon plasma phase  compared with the experiments 
at RHIC and LHC. At this moment, 
we do not have a clear picture under what situation the large $N_c$ approximation are justified\footnote{There are several good examples which work beyond the large $N_c$ and large $\lambda$ limit in the holographic 
setting.  One of the examples
is the Wilson loop, which is nothing but a heavy string trajectory, where we can calculate, by using the localization technique, large $N_c$ but 
any values of $\lambda$ calculation and can see a precise matching between string side and gauge theory side. 
In addition, in the holographic QCD setting,
there is an attempt  even in the large $\lambda$ limit to take into account the degrees of freedom corresponding to the massive open strings whose spins are bigger than 2, and quantize these massive stringy excitation in the weakly curved geometry \cite{Imoto:2010ef}. This also gives a qualitatively very good 
comparison with the experimental data.}.

\subsubsection{ 
Lack of the asymptotic freedom leading to multiple parameters}

QCD is specified by a peculiar energy scale $\Lambda_{\rm QCD}$ as a result of the running coupling constant and
the asymptotic freedom. In particular for the massless QCD, it has 
only this scale in the theory and there is no other parameter. 
In contrast to this, the scale in the holographic QCD is introduced into the system by hand as an input. 
In the holographic QCD, the operators which break the conformal invariance is introduced at some scale $\Lambda_{\rm cutoff}$, so that the coupling constant in the theory becomes scale dependent (the running coupling).

This leads us to a strange situation where we have two scales in holographic system: 
One is the  scale $\Lambda_{\rm cutoff}$ we introduced, and the other is the $\Lambda_{\rm QCD}$.
This $\Lambda_{\rm QCD}$ is an emergent scale in the low energy physics where hadron physics emerges.
This $\Lambda_{\rm QCD}$ is determined as a function of two input parameters, $\Lambda_{\rm cutoff}$ 
and 't Hooft coupling constant $\lambda = g_s N_c$ (where $g_s$ is a string theory 
coupling constant).
In principle, if one 
can take the double-scaling limit at which the $\Lambda_{\rm QCD}$ is fixed while the holographic scale $\Lambda_{\rm cutoff}$ (at which typically particle fields
which do not exist in QCD appear) is taken to infinity, 
by fine-tuning $\lambda$, then the above problem would be resolved.

This, however, is not an easy task:  In order to make the gauge theory coupling constant at the scale $\Lambda_{\rm cutoff}$ to be 
weak,  the corresponding geometry in the holographic side becomes highly curved that the supergravity description is no more reliable. 
As mentioned in the previous section, however, it is technically difficult to go beyond 
large $N_c$ and $\lambda$. As a result, the difficulty of taking the double scaling limit remains in any holographic QCD models in the top-down approach. 

Of course, one can say that the number of the parameters, two given by $\Lambda_{\rm cutoff}$ and $\lambda$, is significantly small, and it is 
good enough to have nontrivial
check and predictions in QCD, compared to many other phenomenological models.

It is noted that 
the coupling constant in the gauge theory becomes strong again beyond $\Lambda_{\rm cutoff}$  due to supersymmetric particles which appear above  $\Lambda_{\rm cutoff}$
\footnote{On the other hand, 
the $\Lambda_{\rm QCD}$ is seen in gravity side as an IR cut-off of the geometry and no geometry exists below that IR cut-off scale (radius).}. 
This property is completely different from the ordinary QCD, where the coupling constant becomes smaller and smaller at higher and higher energy (the asymptotic freedom).

\subsection{Popular holographic models and their problems}
\label{Popularmodel}

Next we shall look at popular holographic models which are widely used for
various purposes, in particular from the viewpoints of their strong points and 
limitations. We make emphasis on the point that, depending on physical quantities of
interest, one can choose a holographic model among many. We here briefly review
five  models popularly used in the top-down approach of the
holographic QCD.

\begin{itemize}
\item Supersymmetric D3-brane model (Asymptotic $AdS_5$)

The gauge-theory counterpart of this model is 
${\cal N}=4$ supersymmetric Yang-Mills theory. This theory is
highly supersymmetric and so is far from the realistic QCD. However, to see robust results of
deconfined gluons in high temperature, where we expect the effect of supersymmetry is not crucial, 
the theory would be sometimes good enough to extract  typical pehnomena
of strong coupling gauge theories. The most successful result which came out of this
is the computation of shear viscosity of quark gluon plasma in high temperature phase
of QCD \cite{viscosity}. Although the computation has employed  only a geometry representing 
a finite temperature phase of the ${\cal N}=4$ supersymmetric gauge theory, the result
is close to the experimental observation. 

It is difficult to argue why this model works so well. 
In terms of the  large $N_c$ expansion, reasons why the $1/N_c$ corrections do not contribute and why they do not modify qualitative nature are still missing.
In addition,
there are many fields in the supersymmetric theory which are absent in QCD. An issue of
the universality of the value of the shear viscosity is still to be settled. 
Nevertheless, other physical quantities have been calculated so far and results give insightful suggestion for heavy ion experiments. 

\item{D3D7 model} 

Introducing D7-branes as flavor D-branes \cite{Karch:2002sh} makes
it possible to include supersymmetric quark fields (hyper multiplets in fundamental representation) in 
above D3-brane model. This make it possible to calculate 
the quark energy loss in quark gluon plasma and drug forces \cite{energylosspaper}. 

One can also discuss $U(1)$ part of chiral symmetry breaking in this D3D7 setting \cite{Babington:2003vm}. 
The position of flavor D7-brane represents the symmetry 
in the Yang-Mills theory on D3-branes. If the position of 
flavor D7-brane are symmetric, we have that symmetry in Yang-Mills theory, however if not, we have 
corresponding symmetry breaking. 
By embedding the $U(1)$ part of chiral symmetry as a geometrical rotational symmetry 
in D3-brane-setting, we can discuss how this rotational symmetry is spontaneously broken from the 
position of D7-brane at low temperature, and restored at high temperature. 
The position of D7-branes is determined in order to minimize the free energy of the system. See for example, 
Fig.~6.2 and Fig.~6.6 of \cite{Erdmenger:2007cm}.

\item{Witten's non-supersymmetric model \cite{Witten:1998zw}}

The corresponding geometry is 
called Gibbons-Maeda geometry \cite{Gibbons:1987ps}, and corresponds to
a  1+3-dimensional  
pure bosonic Yang-Mills theory ({\it i.e.} the theory of gluons) at low energy without supersymmetry. 
The geometry is made of $N_c$ D4-branes wrapping a circle. This circle  compactification
brings the 1+4-dimensional theory down to the 1+3-dimensional Yang-Mills theory. It
breaks the
supersymmetry by imposing anti-periodic boundary conditions for fermions, 
and at low energy all fermions are massive and only massless gluons survive. 
Adjoint scalars obtain masses through quantum corrections which are roughly 
of order of the scale defined by the radius of the circle.

This geometry captures important gauge-theory property: the confinement. 
In fact, using the bulk equations of motion from the gravity side, 
one can demonstrate  as \cite{Witten:1998zw}
that the fluctuation spectrum corresponding to the glueball spectra is discrete and mass-gapped, 
and that the calculated wilson loop shows the area low. 
Furthermore, above a critical temperature 
a phase transition occurs and is interpreted as a confinement-deconfinement transition
since the spectrum becomes continuous at the high temperature phase. 

There is one caveat here: the phase transition scale is nothing but the scale
of the extra circle on which D4-branes wrapping, 
as it is the unique dimensionful physical parameter. So naively speaking,
the higher-dimension cutoff scale is re-interpreted as the scale of 
the gluon theory. The excitation of the massive gluinos and adjoint scalars and their superpartners should 
come into the spectrum above the scale, therefore the theory be no more purely bosonic Yang-Mills theory. 
If one naively ignore those and regard 
the gravity fluctuation as the glueball spectrum made solely of the gluons above this scale, we might 
get some mismatches for the spectrum comparison. 

Therefore the intrinsic problem of this geometry interpreted as a dual of the pure Yang-Mills theory
is the double meaning of the dynamical scale and the compactification scale. 
In order to remove additional degrees of freedom, we have to take the double scaling limit; 
We keep $\Lambda_{QCD} $ fixed and at the same time, take the scale, associated with the D4-brane wrapping circle, to infinity.  
Any proper scaling limit where the dynamical scale is fixed while the compactification scale is taken to infinity, 
has not been formulated yet. 

\item{D4D6 model \cite{Kruczenski:2003uq}}

In the Witten's geometry, flavor D6-branes can be added to include  quarks in the theory.
The string connecting the $N_c$ D4-branes and the $N_f$ D6-branes give a low energy excitation which 
behaves like a quark. In the gravity description, the shape of the D6-branes is deformed
and it can be interpreted as a spontaneous breaking of the (anomalous) $U(1)$ axial symmetry.
The model can include various quark masses, so in particular quark mass dependences
of various low energy quantities can be studied. 

\item{D4D8 model (Sakai-Sugimoto model) \cite{Sakai:2004cn}}

This theory adds flavor D8-branes in the Witten's geometry. One of the superior point of this model 
compared with others is that by adding D8-branes, one can obtain only left-chirality fermions at the intersection points between D4 and D8-branes. On the other hand, by adding anti-D8-brane, one can obtain only right-chirality fermions at the intersection points between D4 and anti-D8 branes. 
This implies that by adding $N_f$ number of both D8 and anti-D8-branes, 
we can have both left and right chiral fermions (quarks) in the system with explicit dependence
on the chiral symmetry $U(N_f)_L\times U(N_f)_R$, 
which are very close to the realistic QCD.

\begin{figure}
\centering
\includegraphics[scale=0.33]{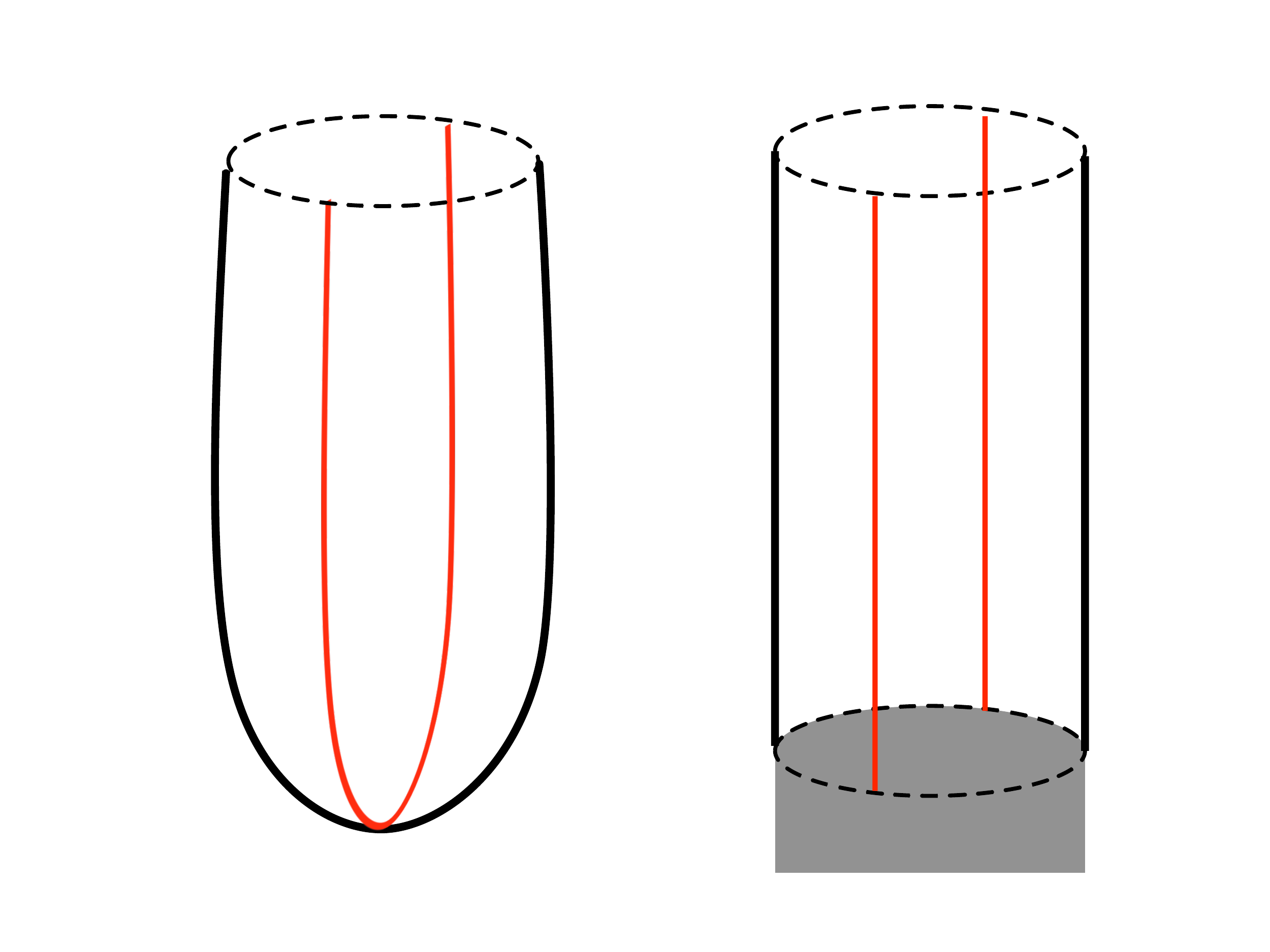}
\caption{
A schematic picture of two phases in the gravity dual of the bosonic pure Yang-Mills theory.
On the geometry specified by the surface of the cylinder, flavor D8-branes (red lines) are put.
The vertical direction in the figure is a holographic dimension, while the circular direction is for the compact circle which brings the 1+4-dimensional theory down to the 1+3-dimensional pure Yang-Mills theory.
The left figure shows a geometry corresponding to a confining phase, while the right one is
that for a deconfined geometry. (Left) The geometry consistently truncated at a certain place along 
the holographic direction corresponding to $\Lambda_{\rm QCD}$, and the D8-brane and the anti-D8-brane are connected due to the geometry,
which shows the spontaneous chiral symmetry breaking. 
(Right) The geometry ends with a horizon of a black hole (shaded region). The D8-brane and the 
anti-D8-brane are independent, which is a chiral symmetry restoration.}
\label{fig4} 
\end{figure}

Similar to the D3D7 system, 
the chiral symmetry $U(N_f)_L\times U(N_f)_R$ is seen from the position of flavor D8 and anti-D8-branes. 
Due to the warped factor of Gibbons-Maeda geometry,  one can demonstrate that 
the free energy at low temperature is lower if both $N_f$ D8-branes and $N_f$ anti-D8-brane are combined into $N_f$ 8-branes. 
 See the left figure of Fig.~\ref{fig4}. 
In high temperature, 
these combined effects of D8 and anti-D8 are hidden behind the horizon  (right figure of
Fig.~\ref{fig4}),
and we have chiral symmetry restoration, which can be seen  
geometrically. 
In this way, this model shows the spontaneous chiral symmetry breaking at low temperature and its restoration 
at high temperature in a geometrical way. 

Except for the point that
the quark mass is difficult to be introduced\footnote{See \cite{Aharony:2008an} for a possible way to introduce the 
quark masses to the model.} due to the non-supersymmetric nature and the existence of chiral matter, 
this holographic model is the most successful
model in view of the study for the low energy hadron physics. In addition to the meson spectrum and interactions,
baryon spectrum and its chiral dynamics can be systematically studied.

This model again suffers from the same problem as the Witten's geometry has: the unnecessary modes,
such as squarks in addition to the gluinos, exist in the theory at high energy scale\footnote{Here holographic scale, which is determined by the scale on which D4-branes are wrapping, gives the scale beyond which these additional ``junk'' are excited.}. So we tentatively ignore modes which are
expected to be absent in QCD, to compare the holographic results with experiments.

\end{itemize}

In summary, in all holographic models popularly known, there remains a problem of
having fields which are absent in real QCD above some scale of the theory. 
And relatedly, the low energy scale of the
theory which one would like to interpret as the QCD scale is shared with the scale
where the unnecessary fields show up. 

Naively, at very low energy, the effects of these unnecessary fields would be small,
so the prediction from holography should be better at the low energy. 
This simple fact would motivate us strongly to visit nuclear physics. 
Nuclear physics treats nuclei: bound states of nucleons
at the energy scale much lower than the QCD scale. 
However in order to make the comparison with data more presice, we have to take the 
double scaling limit in holographic QCD, where we take the scale, beyond which unnecessary fields be dynamical, to infinity while keeping the QCD scale fixed.

Due to the reasons explained in section 1, nuclear physics includes a lot to be explained by QCD.
Standard nuclear physics has many assumptions, 
and the origins of those fundamental assumptions may be
explained directly from QCD, once we apply the holographic methods to QCD.


\section{Review : M(atrix) Description of Multi-Baryon System}
\label{sec3}

The upshot of the theory \cite{Hashimoto:2010je} 
for the multi-baryon system, derived in AdS/CFT, is that it is a theory
of matrix degrees of freedom, with the following robust form of the action:
\begin{eqnarray}
S = \frac{M}{2} \int \! dt \; 
{\rm tr}\left[
\left(\partial_t X^M(t)\right)^2 - g [X^M,X^N]^2 + \cdots
\right]
\label{ourac}
\end{eqnarray}
Let us clarify the relation between \eqref{ourac} and the nuclear physics
action \eqref{nucac}. The matrix $X^M$ is a hermitian $A\times A$ matrix, where $A$ is
the number of baryons (which resultantly becomes the mass number of a nucleus if all the baryons are bounded together as a big nucleus).
Once it is diagonalized, the eigenvalues are nothing but the locations of the baryons
which are given by $x^M_{(s)}$ (for $s=1,\cdots,A$) in the nuclear physics action \eqref{nucac}.
There are off-diagonal entries in $X^M$, which we interpret the degrees of freedom associated with the 
nuclear force mediator (such as pion, massive vector mesons etc). 
Classically integrating those degrees of freedom in the
action gives rise to the interaction between the eigenvalues of $X^M$. For the detail of nuclear force derivations, see  
section 4 of \cite{Hashimoto:2010je}. 
This interaction is interpreted as the inter-nucleon potential (nuclear force).
The terms which are not written in the action \eqref{ourac} (specified as ``$+\cdots$")
are fields representing spins and isospins (flavor degrees of freedom of the baryons).
Again, the precise form of the action is given in \cite{Hashimoto:2010je} and presented in \eqref{mm}.

In this section, we provide a review of the matrix formulation of the multi-baryon system in simple terms.
First, we shall explain below the reason why we have the matrix degrees of freedom for the baryons
in AdS/CFT, and the origin of the action written above. Then we come to a review of the concrete analysis for
a single baryon system to obtain the baryon spectrum, and also a review of two and three baryon 
systems for deriving the short distance nuclear force. These were done in the original paper \cite{Hashimoto:2010je}. 
Then in the final part of this section, we review the importance of the matrix model action \eqref{ourac}
for providing a possible unified view of nuclear physics.

\subsection{Baryons are matrices}

As we outlined above, the most important and novel part of the new description of the multi-baryon system
\eqref{ourac} is the fact that {\it baryons are described by $A\times A$ matrices}. In fact, this is a robust result
once one applies the AdS/CFT correspondence to QCD for the multi-baryon system.

\begin{figure}
\centering
\includegraphics[scale=0.35]{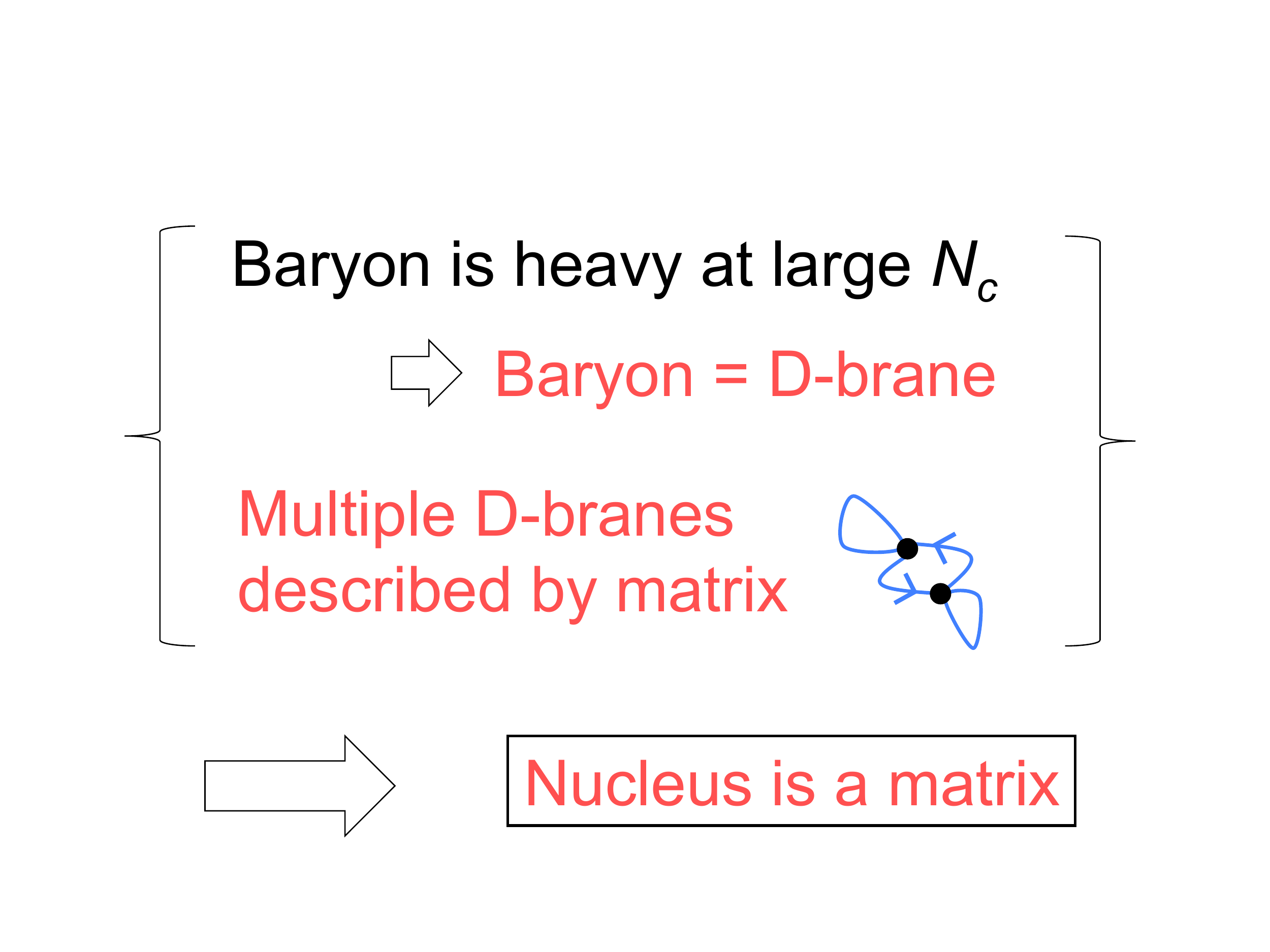}
\caption{Derivation of the matrix description of nuclear physics.}
\label{fig2}
\end{figure}

There are two key points to derive this fact, which are shown in Fig.~\ref{fig2}.
\begin{itemize}
\item {\it A baryon is a D-brane.}

In AdS/CFT correspondence, we need a large $N_c$ expansion to use the dual gravity description.
For large $N_c$ QCD, baryons are heavy object whose mass is of order 
of ${\cal O}(N_c)$, since a single baryon
consists of $N_c$ quarks. In the gravity side of the AdS/CFT, what is the object whose mass is so large?
The answer is D-branes. D-branes are solitonic objects in string theory, whose mass are order 
of ${\cal O}(N_c)$. Therefore the baryons are expected to correspond to
the D-branes in the gravity side of the AdS/CFT correspondence. 
In fact, baryons are D-branes, and technically speaking, these baryon D-branes are wrapping 
on the closed surface like higher dimensional sphere 
on which string theoretic RR flux is penetrating. 
Through the D-brane action, the wrapped D-branes on some closed surface 
with penetrating flux induces $N_c$ unit of $U(1)$ charges on that closed surface. 
On the closed surface, total charges must be zero to satisfy Gauss's law. 
This implies that we need to add compensating charged objects on that surface, which turns out to 
be $N_c$ number of fundamental strings \cite{Witten:1998xy}. Therefore these D-branes behaves 
as baryons.

\item{\it Multi-D-branes are matrices.}

D-branes are defined as surfaces on which open strings can end. 
When D-branes are on top of each other,
fundamental and anti-fundamental strings connecting between those 
D-branes can be arbitrarily short, and can be massless. The low-energy excitation
of those light modes are classified by an $A\times A$ matrix when $A$ is the number of the 
D-branes, since each open string
has two ends labeled as $(a,b)$ where $a,b=1,\cdots,A$. Therefore the low energy 
degrees of freedom on the coincident $A$ 
D-branes are $A\times A$ matrices.

\end{itemize}

Combining these two, we arrive at the inevitable conclusion that nuclei (or the multi-baryon system) 
in the AdS/CFT correspondence should be described by matrices.

Furthermore, the effective action of D-branes has the universal form of \eqref{ourac}. The interpretation is
definite: the eigenvalues of the field $X$ are location of the $A$ number of D-branes. 
Therefore, we come to a conjecture that
the effective action \eqref{ourac} describes nuclear physics.

One of the most important properties of nuclei is its crucial dependence on isospins. 
Nuclear force strongly depends on
whether the nucleon is a proton or a neutron. 
Consequently, we have a nuclear chart and stable/unstable nuclei.
How the isospin dependence can come in in this formulation? The answer is quite simple: another 
matrix $w$ which is an $A \times N_f$ complex matrix 
joins the effective action. The isospins are nothing but the
quark flavor degrees of freedom, and $N_f$ is the number of the quark flavors.

\begin{figure}
\centering
\includegraphics[scale=0.5]{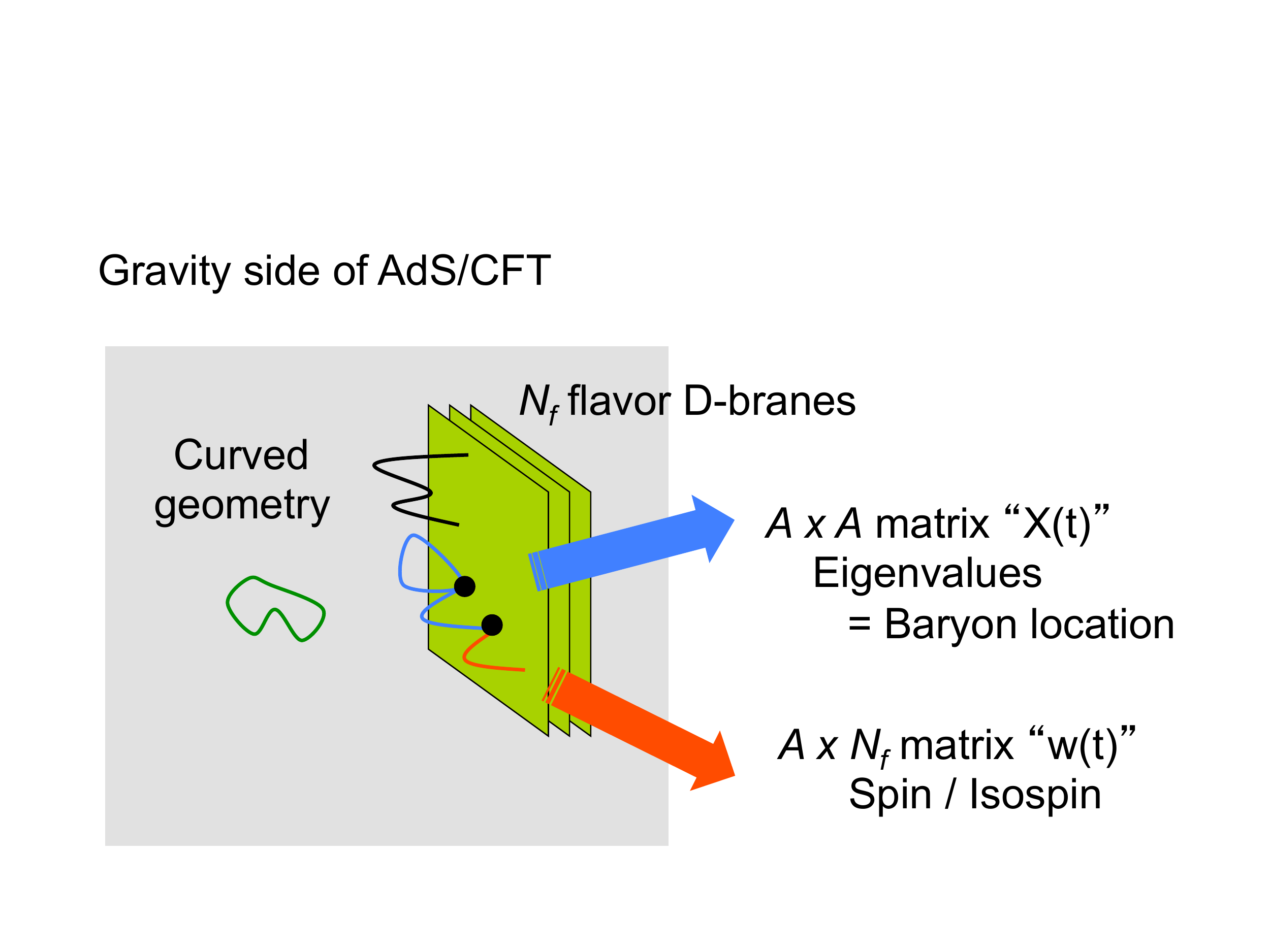}
\caption{The appearance of two kinds of matrices $X$ and $w$. $X$ connects baryon D-branes (depicted as
two black blobs), while $w$ connects a baryon D-brane and a flavor D-brane (depicted as three parallel sheets). $w$ obtains nonzero VEV to satisfy the Gauss's law on each baryon D-brane, which is equivalent to $N_c$ open strings.}
\label{fig3} 
\end{figure}

Fig.~\ref{fig3} clarifies why this new matrix shows up in the gravity side of the AdS/CFT correspondence.
As we reviewed in the previous section, the flavor can be represented by an 
introduction of ``flavor D-branes"
into the gravity geometry. Then, in addition to the baryon D-branes, we have the flavor D-branes, so
there appears an open string which connects the two kinds of D-branes. This string should be described
by $A\times N_f$ matrices, as in the same manner as the $A\times A$ 
matrix $X$ for the string among the
baryon D-branes.

Although the species of the fields appearing in the low energy of the multi-baryon system in the 
AdS/CFT are just $X$ and $w$, the precise interaction between these fields, and also the coefficients
in the effective action, depends on what kind of D-brane configurations (holographic models) we use 
for the large $N_c$ QCD. When we use the most popular D4D8 model (Sakai-Sugimoto model) described 
in section \ref{Popularmodel}, the baryon 
D-branes are D4-branes wrapping $S^4$, and the flavor D-branes are D8-branes wrapping $S^4$. 
This means that the baryon D4-branes can be located inside the flavor D8-branes. 
The D$p$-D$(p+4)$ system in superstring theory is well-understood, 
as a geometric realization of the instanton construction: 
the D$p$-brane can be seen as a Yang-Mills instanton through the gauge fields on the D$(p+4)$-brane,
where the dynamics of the D$p$-brane can be determined by a so-called ADHM matrices 
used for the instanton construction \cite{Atiyah:1978ri,Dorey:2002ik}.
Therefore, within the D4D8 holographic model, 
our low energy effective action for the multi-baryon system is
nothing but a generalization of the ADHM matrix models.  

The matrix effective action is concretely written in 
\eqref{mm} for D4D8 model, but in this review part we don't need the explicit form,
as we explain only the conceptual part to show the robustness of the derivation. 
Furthermore, it is straightforward to construct explicit forms for the matrix effective action for 
another D-brane models described in subsection \ref{Popularmodel}. However for concreteness, 
in this paper we concentrate on the model constructed for D4D8 model in \cite{Hashimoto:2010je}. 

Next, we give a review of 
a single baryon spectrum $(A=1)$, and also a derivation of the short distance nuclear force $(A=2,3)$.
The important fact for the application is that the matrix action has only two free parameters:

\subsection{Derived baryon spectrum}

The simplest case is $A=1$ where we have only a single baryon. In this case, the quantum mechanics
should give the baryon spectrum. Excited states of a baryon emerges from the quantum mechanics.

Let us recall the Skyrme model \cite{Skyrme:1962vh,Adkins:1983ya}. 
In the Skyrme model, a baryon appears as a soliton of the Skyrme model
which is nothing but a peculiar effective action of low energy pions. Any soliton has fluctuation modes,
massive or massless (zero modes). The fluctuation modes, which are just a function of time, obey a 
hamiltonian, and they can be quantized. The resulting quantized fluctuation spectrum is interpreted
in the Skyrme model as the baryon spectrum. Here in the AdS/CFT matrix model 
approach, the hamiltonian of the fluctuation 
modes are directly given as our matrix model hamiltonian \eqref{ourac}. So, one easy interpretation
of the matrix model is a moduli hamiltonian of generalized Skyrmions. 
However very small number of parameters (only two parameters) in our holographic setting gives the superiority 
of our construction compared with generic Skyrme model which have many parameters. 

For $A=1$, the matrix model becomes extremely simple. 
The hamiltonian for two flavors ($N_f=2$) looks \cite{Hashimoto:2010je}
\begin{eqnarray}
H = \frac{\lambda N_c M_{\rm KK}}{54\pi}
\left[
\left(\frac{27\pi}{\lambda M_{\rm KK}}\right)^2 \frac{1}{2\rho^2}
+ \frac{1}{3}M_{\rm KK}^2 \rho^2 + \frac{2}{3}M_{\rm KK}^2 (X^4)^2
\right],
\end{eqnarray}
where 
\begin{eqnarray}
w^{i}_{\dot{\alpha}} = \rho(t) U_{\dot{\alpha}}^i(t).
\end{eqnarray}
Here $\rho(t)$ and $X^4(t)$ are scalar degrees of freedom, 
and $U(t)$ is a $2\times2$ unitary matrix degree of freedom. $\rho (t)$ represents dissolved 
size of the D-brane (which is roughly the size of the baryons), and $X^4(t)$ 
is a displacement of the D-brane along the holographic direction. The matrix $U$ is
nothing but the moduli degrees of freedom appearing in the Skyrme model,
and its quantization gives higher spins and isospins.

\begin{figure}
\centering
\includegraphics[scale=0.4]{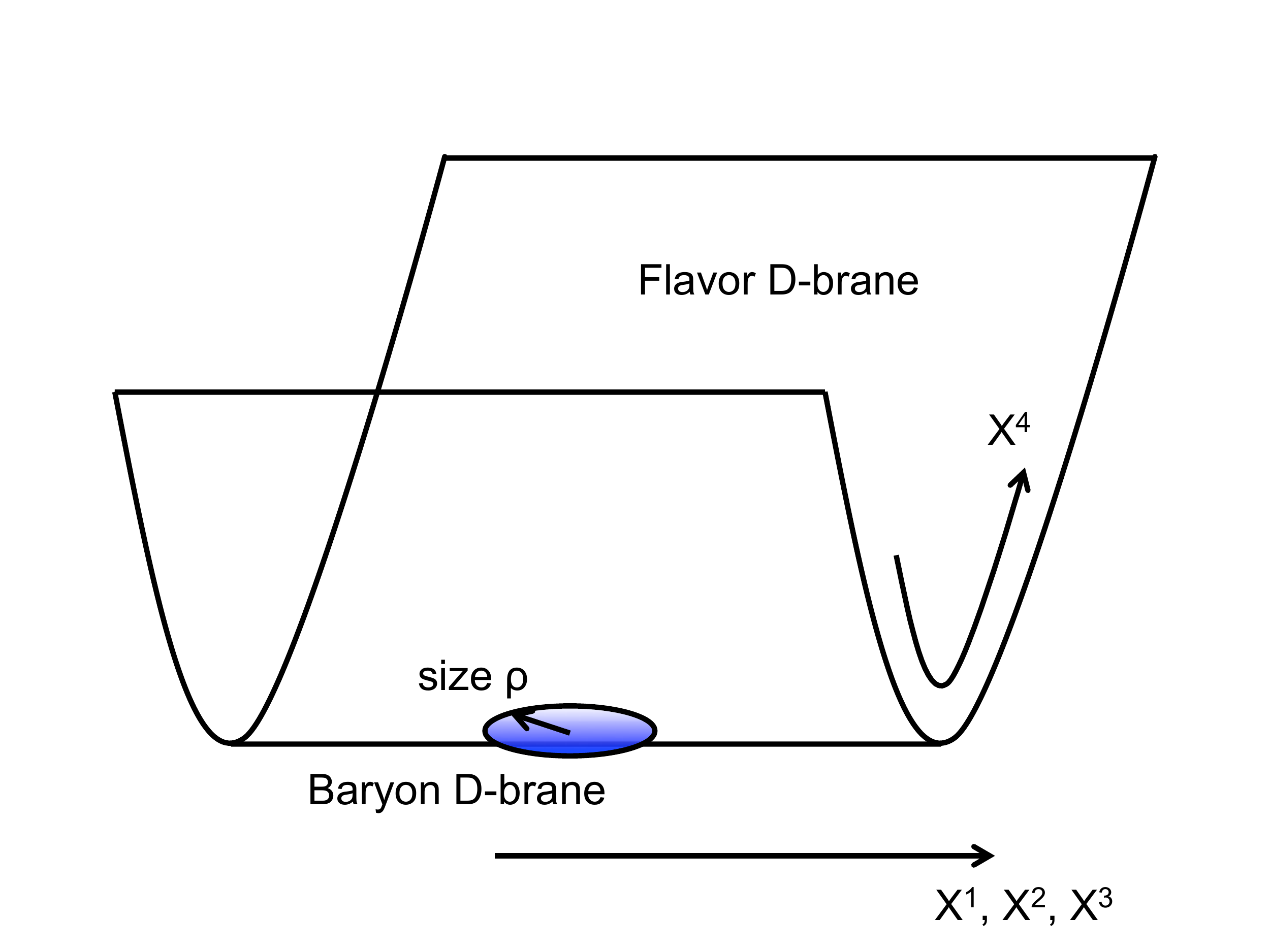}
\caption{A schematic picture of the baryon D-brane on the flavor D-brane. The harmonic oscillator 
excitation
on the baryon D-brane is the size fluctuation $\rho$ and the fluctuation of the D-brane location 
along the holographic direction $X^4$. The horizontal direction is our space $x^1,x^2,x^3$.}
\label{figbaryon}
\end{figure}

These three modes provides almost-independent harmonic oscillators,
and the quantization results in the following spectra:
\begin{eqnarray}
M = M_0 + \frac{M_{\rm KK}}{\sqrt{6}}
\left[\sqrt{(I/2+1)^2+ N_c^2}+ 2n_\rho + 2n_{X^4} + 2
\right].
\label{spec}
\end{eqnarray}
Here $I$ is the isospin which is equal to the spin in the present case, and $n_\rho$ 
$(n_{X^4})$
is a non-negative integer coming from the harmonic oscillator $\rho(t)$ $(X^4(t))$.

The baryon spectrum \eqref{spec}, as well as its calculation from the quantum hamiltonian,
is quite close to what has been obtained in the soliton quantization approach 
in the Sakai-Sugimoto model \cite{Hata:2007mb}. 

We again would like to stress that the result is qualitatively robust in the AdS/CFT approach:
because the baryon in the gravity dual should be represented by $X$ and $w$ strings,
the spectrum should be given by its low energy quantization. So we are inevitably led to
the quantum number $n_{X^4}$ which is the oscillation of the D-brane along the holographic
directions, and also the quantum number $n_\rho$ which is the fluctuation of the magnitude
of the string connecting the baryon D-brane and the flavor D-brane, and also the spin operator
$U$ which is the internal orientation of the same string. The coefficients appearing in the
mass spectrum formula may differ among holographic models, but its structure
should be shared in all the holographic models.

\subsection{Universal repulsive core of nucleons}

Once the baryon state can be identified within the matrix model degrees of freedom, it
is straightforward to calculate the inter-baryon potential. Since the short-distance behavior
of the nuclear force is one of the most important problems in nuclear physics,
to derive it analytically is a very important issue. As we have the matrix model action
for $A=2$ at hand, whether it reproduces the empirically-known repulsive core of nucleons
would be a good touchstone for the validity of the matrix model approach.

Since the matrix model action \eqref{ourac} is explicitly given for $A=2$,
we just need to: first derive the off-diagonal term classically by solving ADHM constraints\footnote{For detail, see section 4.1.2. of \cite{Hashimoto:2010je}.} for a given set of two diagonal
entries which defines the locations and the spin/isospins of the two baryons, and then substitute all 
back into the hamiltonian to derive the inter-baryon potential energy.

The calculation is straightforward and was given in \cite{Hashimoto:2010je}.
The result for the inter-nucleon potential is as follows:\footnote{
Using the soliton approach in the holographic D4D8 model, the short distance nuclear force
was calculated \cite{Hashimoto:2009ys}. The result is qualitatively similar to the result of the matrix model.
}
\begin{eqnarray}
&&V_{\rm central}(r)
= \frac{\pi N_c}{\lambda M_{\rm KK}} \left(
\frac{27}{2} + 8 \vec{I}_1 \cdot \vec{I}_2 \; \vec{J}_1\cdot \vec{J}_2
\right) \frac{1}{r^2}, \quad
\\
&& V_{\rm tensor}(r)
= \frac{2\pi N_c}{\lambda M_{\rm KK}} \vec{I}_1 \cdot \vec{I}_2 \; \frac{1}{r^2}.
\end{eqnarray}
This short-distance potential is positive for any choice of the spins and the isospins, therefore 
we conclude for $N_f = 2$ case, there are universal repulsive cores for the nuclear forces at short 
distance. This repulsive core behaves as  $1/r^2$, showing very strong repulsive core at $r \to 0$ limit.
We concluded that qualitatively the matrix model approach for multi-baryon
system is consistent with experiments, in this sense. Whether the repulsive core behaves as 
$1/r^2$ or not should be tested in future. 

The three-body nuclear force can be evaluated in the same manner. 
The short distance contribution to the intrinsic three-body force, 
which does not come from the effective integration of
massive states (for example the famous Fujita-Miyazawa force \cite{Fujita:1957zz}), is
important as it cannot be evaluated using chiral perturbations. Using the
matrix model approach, one can straightforwardly evaluate 
the three-body interaction. It was shown in \cite{Hashimoto:2010ue} that the proton-proton-neutron
aligned on a line gives a positive three-body potential, and that a spin-averaged three-neutron
aligned on a line is positive too. These are consistent with experiments\footnote{See also the recent attempt in lattice QCD \cite{Doi:2011nm}.}. In particular, the latter
is relevant for neutron stars as it has a dense neutron system, and the effective repulsion
would give a hard equation of state which is a good tendency for a recent observation of heavy
neutron stars.


\subsection{Toward a description of atomic nuclei}

As we have the effective action \eqref{ourac} of the multi-baryon system, in principle,
atomic nuclei and their properties, {\it i.e.} the nuclear physics, should emerge from the action.
The action \eqref{ourac} has only two parameters, so once one can solve the 
effective action completely in a quantum mechanical fashion, one can compare the results with experiments in principle. Whether this action provides us with an efficient and good description of atomic nuclei
is a very important question, as the AdS/CFT connects directly the nuclear physics and QCD.

In particular, properties of heavy nuclei are yet to be uncovered, and they are far from QCD.
The aim of the holographic approach is to uncover the relation between the nuclear physics
and QCD directly, to make clear how observables in nuclear physics may depend on
quantities defined in QCD. One of the important targets in nuclear physics in this sense is
the nuclear radius. It has been known for many decades that stable nuclei are subject to
a relation 
\begin{eqnarray}
r\sim 1.2 \times A^{1/3} \; \mbox{ [fm] }
\label{nr}
\end{eqnarray}
where $A$ is the mass number (the number of baryons)
of the nucleus. This has been explained as a result of the nuclear density saturation:
the nucleon density inside nuclei is almost constant and takes a universal value, so the 
nuclear radius is proportional to $A^{1/3}$.

The repulsive core of nucleons is thought to be a component to explain the $A$ dependence of the
nuclear radius in nuclear physics. If a nucleon can be regarded as a hard ball which is almost equivalent to
the repulsive core, the total nucleus should have a volume proportional to $A$, therefore the
$A$ dependence follows. Since in the holographic QCD approach the repulsive core was reproduced
as explained in the previous subsection, this nuclear radius would be a natural consequence.

In \cite{arXiv:1103.5688}, one of the authors (K.H.) together with T.~Morita demonstrated 
that indeed the above \eqref{nr} is reproduced 
from the quantum mechanical matrix action \eqref{ourac}, with a certain approximation
employed. The result obtained in \cite{arXiv:1103.5688} is an analytic formula for the
nuclear radius,
\begin{eqnarray}
\sqrt{r^2_{\rm mean}}
=
\frac{3^{5/2}\pi^{2/3}}{2^{5/6}5^{1/6}} \frac{1}{M_{\rm KK} \lambda^{2/3} N_c^{1/3}} \, A^{1/3}.
\label{nrc}
\end{eqnarray}
A nontrivial point is that the formula has the correct $A^{1/3}$ dependence.

\begin{figure}
\centering
\includegraphics[scale=0.2]{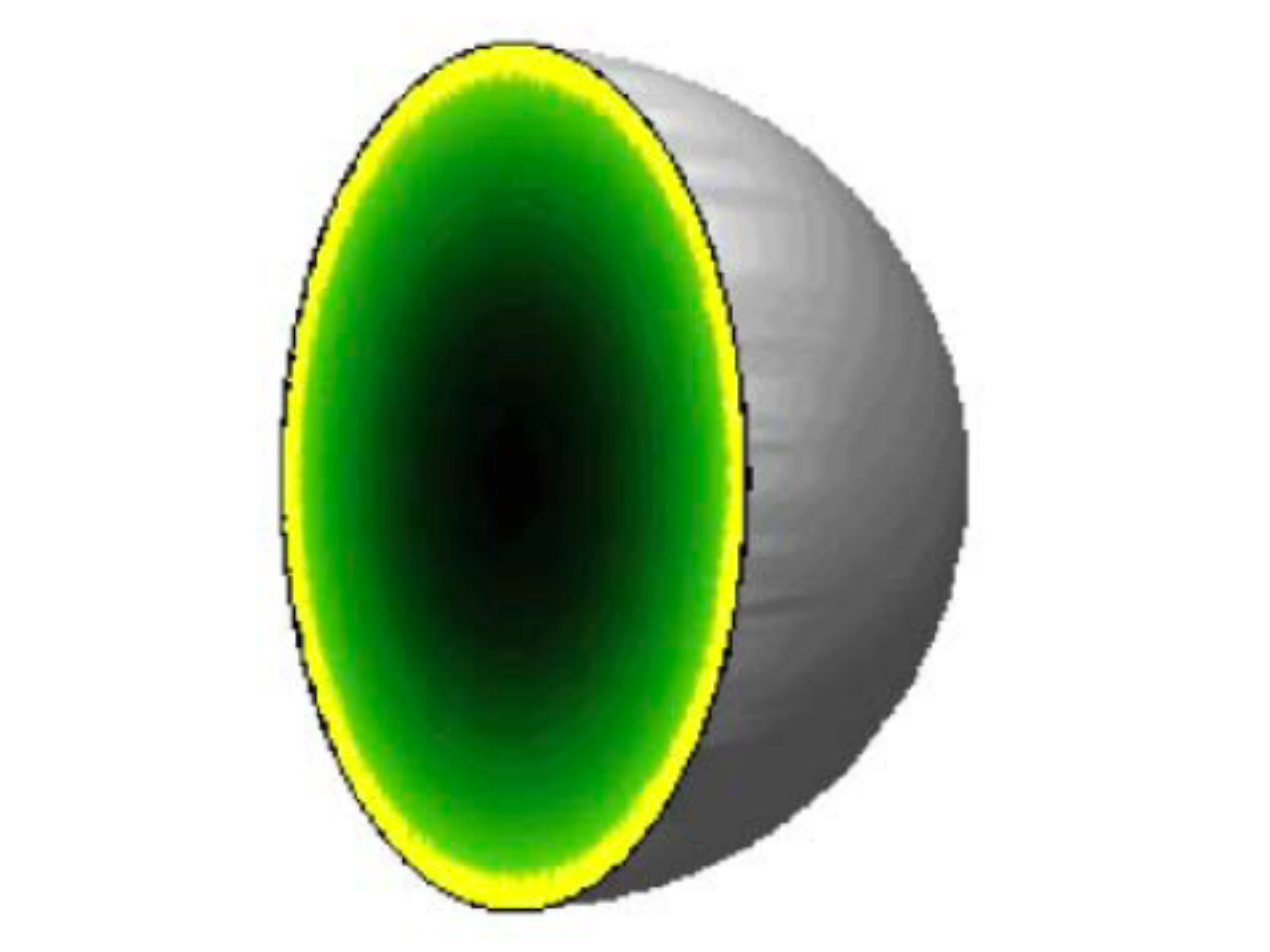}
\caption{A plot of a nuclear density inside the nucleus, calculated in AdS/CFT \cite{arXiv:1103.5688}. }
\end{figure}

The approximation used for deriving \eqref{nrc} is: a large $A$ limit with quenching (ignoring 
the $w$ degrees of freedom), and a large dimension limit (which is almost equivalent to 
a mean field limit), in addition to the standard limits taken in holographic QCD, such as
the large $N_c$ limit and the strong coupling limit $\lambda \gg 1$.
Whether these approximations are appropriate or not should be studied in the future study.
However, the message here is that we now have a good starting point \eqref{ourac}
for calculating various quantities in nuclear physics from QCD. 

It would be possible that the action \eqref{ourac} itself may be modified to include
higher order terms, or that one needs to apply different approximations to \eqref{ourac}
to correctly derive physical observables from \eqref{ourac}. 
Especially higher order terms are crucial when we consider the effect of large $A$ limit. 
The original action \eqref{ourac} is derived at the limit where baryon D-branes are at the bottom of 
the warped geometry, see Fig.~\ref{figbaryon}. As the size be bigger, we need to consider the effect of 
curved geometry more precisely and this gives the higher derivative corrections to the 
action \eqref{ourac}\footnote{In addition, one may take a gravity dual of the matrix model under a certain assumption
(such that $A$ is large and also that the inter-nucleon distance is small) to
investigate giant resonances in nuclei \cite{Hashimoto:2008jq}.}.
All of those efforts can
be a good bridge between QCD and nuclear physics.


\section{Multi-Flavor Nuclear Forces Via Holography}

\subsection{Strangeness and holography} 

QCD at high density is a final frontier,
which is still remains to be unveiled. It is expected that at the core of neutron stars,
high density matter would be supplemented with strange quarks, in order to
relax the Fermi energy of the ordinary two-flavor matter of neutrons and protons. 
To judge whether the strangeness really kicks in to the high density core of the neutron
stars, we need to know inter-baryon interaction with multi-flavors. It has been known
experimentally for many decades that nucleons are accompanied with repulsive cores, the short 
distance repulsion. To reveal whether there exists the repulsive core even for baryons
including strange quark(s) is an indispensable cornerstone to reach the truth in
the high density QCD.

The inter-baryon potential is a non-perturbative regime of QCD, even at the short distances
of concern. Thus we need to rely on analytic method to solve the QCD approximately. 
In this section, 
we shall
extend the analysis \cite{Hashimoto:2010je}
to the case of multiple flavors $N_f>2$, to find the short distance properties of the inter-baryon
potential.

Another non-perturbative framework of QCD, lattice simulations, recently uncovered 
interesting features of the short-distance inter-baryon potential for the case of three flavors. 
It was demonstrated \cite{Inoue:2010hs,Inoue:2010es} that indeed 
there remains a repulsive core, so the repulsion is universal. 
\footnote{There is a channel at which the repulsive core disappears,
for an appropriate choice of the baryon states. It
is closely related to the conjectured two-baryon
bound state called H-dibaryon  \cite{Jaffe:1976yi}, as demonstrated in lattice simulations
\cite{Inoue:2010es,Beane:2010hg}. 
Whether such dibaryon exists or not should be confirmed by future experiments.}
In this section, we find that 
the holographic QCD shows the universal repulsive core for generic states in
multi flavors.


\subsection{The effective model of multi-baryon system}

To extract the non-perturbative potential among baryons
at short distances, the nuclear
matrix model \cite{Hashimoto:2010je} derived in holographic QCD
should provide a good sense of the generic nature. 
The action of the model is a quantum mechanics,
\begin{eqnarray}
S &=& 
\frac{\lambda N_c M_{\rm KK}}{54 \pi} 
\int\! dt \; {\rm tr}_k 
\left[
(D_0 X^M)^2 -\frac23 M_{\rm KK}^2 (X^4)^2
\right.
\nonumber \\
&&\left.
+ \, D_0 \bar{w}^{\dot{\alpha}}_i D_0 w_{\dot{\alpha}i} 
- \frac16  M_{\rm KK}^2 \bar{w}^{\dot{\alpha}}_i w_{\dot{\alpha}i}
+ \frac{3^6 \pi^2}{4 \lambda^2 M_{\rm KK}^4} 
\left(\vec{D}\right)^2
\right.
\nonumber \\
&&\left.
+ \, \vec{D}\cdot \vec{\tau}^{\;\dot{\alpha}}_{\;\;\;\dot{\beta}}
\bar{X}^{\dot{\beta}\alpha} X_{\alpha \dot{\alpha}}
+  \vec{D}\cdot \vec{\tau}^{\;\dot{\alpha}}_{\;\;\;\dot{\beta}}
\bar{w}^{\dot{\beta}}_{i} w_{\dot{\alpha}i}
\right]
\nonumber \\
&&
+  \, N_c \int \! dt\; {\rm tr}_k A_0 \, . 
\label{mm}
\end{eqnarray}
The system possesses a gauge symmetry $U(k)$ where $k$ is the
number of the baryons in the system. 
The table 1 shows the field content of the model.
\begin{figure}
\begin{center}
\begin{tabular}{|c|c||c|c|c|}
\hline
 field & index & $U(k)$ & $SU(N_f)$ & $SU(2)\times SU(2)$\\\hline
\hline
$X^M(t)$ & $M=1,2,3,4$ & adj. & ${\bf 1}$& $({\bf 2}, {\bf 2})$\\\hline
$w_{\dot{\alpha}i}(t)$ & $\dot{\alpha}=1,2$; $i=1,\cdots,N_f$
& ${\bf k}$ &${\bf N_f}$ & $({\bf 1},{\bf 2})$  \\\hline
$A_0(t)$ & & adj. & ${\bf 1}$ & $({\bf 1}, {\bf 1})$ \\\hline
$D_s(t)$& $s=1,2,3$ & adj. & ${\bf 1}$& $({\bf 1},{\bf 3})$\\\hline
\end{tabular}
\caption{Fields in the nuclear matrix model.}
\end{center}
\end{figure}

Here, the dynamical fields are $X^M$ and $w_{\dot{\alpha}i}$, while $A_0$
and $D_s$ are auxiliary fields. In writing these fields, the indices for
the gauge group $U(k)$ are implicit. 
The symmetry of this matrix quantum mechanics is
$U(k)_{\rm local}\times SU(N_f)\times SO(3)$ 
where the last factor $SO(3)$ is the spatial rotation, which, together
with a holographic dimension, forms a broken
$SO(4)\simeq SU(2)\times SU(2)$ shown in the table. The breaking is due
to the mass terms for $X^4$ and $w_{\dot{\alpha}i}$.  
In the action, the trace is over these $U(k)$ indices, and 
the definition of the covariant derivatives is
$D_0 X^M \equiv \partial_0 X^M -i[A_0, X^M]$,
$D_0 w \equiv \partial_0 w -i w A_0$ and 
$D_0 \bar w \equiv \partial_0 \bar w + i A_0 \bar w$.   
The spinor indices of $X$ are defined as
$X_{\alpha\dot{\alpha}}\equiv X^M(\sigma_M)_{\alpha\dot{\alpha}}$ and
$\bar{X}^{\dot{\alpha}\alpha}
\equiv X^M(\bar{\sigma}_M)^{\dot{\alpha}\alpha}$
where $\sigma_M=(i\vec{\tau}, 1)$ and
$\bar{\sigma}_M = (-i\vec{\tau},1)$, with Pauli matrices $\tau$.
The model has a unique scale 
$M_{\rm KK}$, and $\lambda = N_c g_{\rm QCD}^2$ 
is the 'tHooft coupling constant of QCD, with the number of colors $N_c$. 
The diagonal entries of $X^i$ $(i=1,2,3)$ specify the location of the baryons.
The location of the baryon D-brane in the holographic direction $X^4$ is stabilized at $X^4=0$ 
around which the harmonic excitations label excited baryon states. The $w$ fields are responsible
for spins and isospins (and flavor representations) of each baryon.

In \cite{Hashimoto:2010je}, explicitly demonstrated is the 
nuclear force for the two-flavor case. There, a universal repulsive core was found.
We here simply extend the two-flavor calculation
to the case with a generic number of massless flavors, and will see the consequence.

The procedure we employ in the following is as follows. First, we look at the
configuration which minimizes the potential of the matrix model. When taking a large $\lambda$,
the D-term condition is required to be satisfied, which is nothing but the ADHM constraint.
Then, we obtain a classical potential with a solution of the ADHM constraint, which depends on
the inter-baryon distance and the moduli parameters of the two baryons.
Taking an expectation value of this potential with respect to the product of the wave functions
for each baryon, we obtain the inter-baryon potential.


\subsection{Two-baryon configuration}

The large $\lambda$ limit lets only configurations satisfying the ADHM constraint
remain. The ADHM constraint is equivalent to the D-term condition concerning $D_s$, and
is given by
\begin{eqnarray}
\vec{\tau}^{\;\dot{\alpha}}_{\;\;\;\dot{\beta}}
\left(
\bar{X}^{\dot{\beta}\alpha} X_{\alpha\dot{\alpha}} + \bar{w}^{\dot{\beta}}_{\;\;\; i} w_{i\dot{\alpha}}
\right)_{BA}=0.
\label{ADHM}
\end{eqnarray}
Here $A,B$ are $U(k)$ indices. For a single baryon with generic number of flavors, 
the ADHM constraint is simply solved by $X=$ constant (baryons located anywhere), and 
\begin{eqnarray}
w= \; U 
\left(
\begin{array}{cc}
\rho& 0\\
0 & \rho\\
0 & 0 \\
\cdots & \cdots \\
0 & 0
\end{array}
\right),
\label{singlebaryon}
\end{eqnarray}
which shows the $(i,\dot{\alpha})$ entry. 
Here $U$ is a $U(N_f)$ unitary matrix specifying the baryon spin and isospin (flavor dependence).
The flavor symmetry acts on $U$ as $U \mapsto GU$.
The baryon wave function is given as $\psi(U)$, as in the same manner as the famous Skyrme model.

We want to put two baryons located at $x_M=\pm r_M/2$, so that the distance between the
two baryons is $r_M$. For the two baryons, now the coordinate field $X_M$ is two by two matrices,
so we parameterize them as 
\begin{eqnarray}
X_M = \frac12 r_M^a \tau_a
\end{eqnarray}
where $\tau_a$ is a Pauli matrix with index $(A,B)$. We specify the baryon location by the diagonal
entries $r^3_M$, while assuming the off-diagonal $r^1$ and $r^2$ are small as $1/(r^3)$, so that
the distance defined by $r_3$ makes sense at large $r^3$. The two baryons can have independent spins and
flavor representations, so we allow
\begin{eqnarray}
w^{A}=  U^{(A)} 
\left(
\begin{array}{cc}
\rho_A & \!\!\!0\\
0 & \!\!\!\rho_A\\
0 & \!\!\!0 \\
\cdots & \!\!\!\cdots \\
0 & \!\!\!0
\end{array}
\right)
\left({\bf 1}_{2\times 2} + \epsilon^{(A)}\right)
,
\nonumber
\end{eqnarray}
for each baryon, $A=1,2$. (In this expression we don't make a summation over
the index $A$.) And $\epsilon^{(A)}$ is taken to be a $2\times 2$ traceless matrix
at ${\cal O} (1/(r^3)^2)$. 
Note that 
at the large inter-baryon distance limit $r^3\rightarrow \infty$, the ADHM data
above reduces to just a set of two single-baryon ADHM data, \eqref{singlebaryon}
and a constant diagonal $X$.

It is quite straightforward to solve the ADHM constraint \eqref{ADHM} with the above
generic ansatz, and the solution is given as follows.
\begin{eqnarray}
&& r^1_M\sigma_M =
\frac{-\rho_1\rho_2}{|r^3|^2} r^3_M \sigma_M  (P_{12}-P_{12}^\dagger) ,
\label{r1sol}
\\
&& 
 r^2_M\sigma_M =
\frac{-i\rho_1\rho_2 }{|r^3|^2} r^3_M \sigma_M  (P_{12}+P_{12}^\dagger),
\label{r2sol}
\\
&& \epsilon^{(1)}=
\frac{-\rho_2^2}{4 |r^3|^2} [P_{12},P_{12}^\dagger] 
, \quad
\epsilon^{(2)}=
\frac{\rho_1^2}{4 |r^3|^2} [P_{12},P_{12}^\dagger] . \quad
\end{eqnarray}
Here we have defined
\begin{eqnarray}
P_{12} \equiv {\bf P}
\left[
(U^{(1)})^\dagger U^{(2)}
\right], 
\end{eqnarray}
with ${\bf P}$ being a projection of the $N_f\times N_f$ matrix to its
upper-left $2\times 2$ components, so that $P_{12}$ is a $2\times 2$
matrix. We can easily see that, when $N_f=2$, the result here can reproduces
the two-flavor result of \cite{Hashimoto:2010je}.


\subsection{Explicit inter-baryon potential}

Let us substitute the above ADHM data, the two-baryon configuration with the
inter-baryon distance $r^3_M$ and the spin/flavor dependence $U^{(A)}$, 
into the action \eqref{mm} and derive the inter-baryon potential as a function
of $r^3$ and $U^{(A)}$. As was done in \cite{Hashimoto:2010je}, we need to integrate out
the $U(2)$ auxiliary gauge field $A_0$ of the quantum mechanics,
\begin{eqnarray}
A_0 = A_0^0 {\bf 1}_{2\times 2} + A_0^a \tau^a.
\end{eqnarray}
Since the model includes
only the linear and quadratic terms in $A_0$, it is straightforward to perform the integration. 
In the action \eqref{mm}, the terms relevant to $A_0$ are
\begin{eqnarray}
S_{A_0} = \frac{\lambda N_c M_{\rm KK}}{54 \pi}
\int \! dt 
\biggl[
2(A_0^a)^2 (r_M^b)^2 - 2(A_0^a r_M^b)^2
\biggr.
\nonumber \\
+ \left(
(A_0^0)^2+ A_0^a)^2
\right) \left(|w^{A=1}|^2 + |w^{A=2}|^2\right)
\nonumber \\
+ 2
A_0^0A_0^1
 \left(w^{A=1}\bar{w}^{A=2} + w^{A=2}\bar{w}^{A=1}\right)
  \nonumber \\
-2i
A_0^0A_0^2
 \left(w^{A=1}\bar{w}^{A=2} -w^{A=2}\bar{w}^{A=1}\right)
\nonumber \\
\biggl.
+ 2
A_0^0A_0^3
 \left(|w^{A=1}|^2 - |w^{A=2}|^2\right)
+ \frac{108\pi}{\lambda M_{\rm KK} }A_0^0
\biggr].
\end{eqnarray}
We integrate out all the components\footnote{Note that \eqref{r1sol} and \eqref{r2sol} satisfy 
$r^1_M r^3_M = r^2_M r^3_M=0$
which may help reducing the $A_0$ action.} of the auxiliary field $A_0$ and write the potential as
$S_{A_0} = - \int dt \; V_{A_0, {\rm 2-body}}$. We expand the result in terms of small $\rho/r^3$ (note that
$r^3$ is the distance between the baryons in $x^3$ direction, not the cubic power of $r$!), to obtain the leading
term
\begin{eqnarray}
V_{A_0, {\rm 2-body}}
= \frac{27\pi}{4}\frac{N_c}{\lambda M_{\rm KK}}
\frac{1}{(r^3)^2}
\bigl|{\rm tr} P_{12}
\bigr|^2.
\label{pota0}
\end{eqnarray}
Here we have already put $\rho_1=\rho_2 =\rho$ which is ensured at large $N_c$.

The remaining contributions to the inter-baryon potential, from the matrix model action,
is the mass terms ${\rm tr} (X_4^2)$ and $|w|^2$. Substituting the two-baryon
configuration, we obtain
\begin{eqnarray}
&&V_{X^4, {\rm 2-body}} 
 =  \frac{\lambda N_c M_{\rm KK}^3}{162 \pi}
\frac{-\rho_1^2\rho_2^2}{((r^3)^2)^2}
\hspace{10mm}
\nonumber
\\
&&
\hspace{10mm}
\times {\rm tr}\left[r^3_M \sigma_M P_{12}^\dagger  
\right]
{\rm tr}\left[r^3_M \sigma_M P_{12}  
\right].
\label{potx4}
\end{eqnarray}
It turns out that the mass term for $w$ does not give rise to an extra potential.

So, in total, the inter-baryon potential $V$ in the small $\rho/r^3$ expansion is given as a sum of \eqref{pota0} and
\eqref{potx4}, 
\begin{eqnarray}
V_{{\rm 2-body}}
= \frac{27 \pi N_c}{4 \lambda M_{\rm KK}}
\frac{1}{|\vec{r}|^2}
\left(
\bigl|{\rm tr}P_{12}\bigr|^2 + 
 \bigl| {\rm tr}\bigl[\vec{\hat{r}}\cdot\vec{\tau} \;P_{12}\bigr]\bigr|^2
\right). \hspace{2mm}
\label{pot}
\end{eqnarray}
Here, we already substituted $r^3_{M=4}=0$ which is satisfied by the baryon wave functions
at large $N_c$ \cite{Hashimoto:2010je},
and denoted $r^3_{M=1,2,3}$ as $\vec{r}$, the inter-baryon vector. $\vec{\hat{r}}$ is the
unit vector along $\vec{r}$, and we also used the classical size $\rho$ of a 
single baryon \cite{Hashimoto:2010je}, $
\rho_1^2 = \rho_2^2 = 3^{7/2} \pi/(\sqrt{2}\lambda M_{\rm KK}^2)$.

We immediately notice that by taking $N_f=2$ the potential \eqref{pot} reduces that of
the two-flavor inter-nucleon potential given in \cite{Hashimoto:2010je}. 
The potential has the $1/|\vec{r}|^2$ behavior which is peculiar to the holographic QCD 
\cite{Hashimoto:2009ys,Hashimoto:2008zw,Kim:2009sr},
which is nothing but a harmonic potential in 4-dimensional space (our spatial 
3 dimensions plus the holographic direction).


\subsection{Universal repulsive core}

It is already manifest that the inter-baryon potential for generic number of flavors, \eqref{pot}, is
positive-semi-definite, since \eqref{pot} is a sum of two positive semi-definite terms. Therefore,
we conclude that holographic QCD predicts positive-semi-definite repulsive core for combination of 
any two baryon states.

Looking at the magnitude of the potential, we notice the following important fact: 
As the number of the flavors is larger than 3, the classical potential \eqref{pot} can vanish,
for appropriate choice of the baryon state. This is simply because we can choose a set of the unitary matrices
$U^{(1)}$ and $U^{(2)}$ such that ${\bf P}((U^{(2)})^\dagger U^{(1)})$ vanishes. As the projection
operator ${\bf P}$ refers only the upper-left corner of the unitary matrices, once the size of the
matrix $N_f$ gets larger, the configuration of the baryon can evade the upper-left $2\times 2$
corner, and thus does not contribute to the inter-baryon potential \eqref{pot}.

Substituting some particular values of constant $U$ corresponds to a classical evaluation of the potential
(as in the same manner as the Skyrme model), but in reality we need to take into account the baryon wave
function $\psi_1(U^{(1)}) \psi_2(U^{(2)})$. A generic wave function has a wide distribution over the space
of the unitary matrices normalized.
So the magnitude of the repulsive core depends on the two baryon states.
The situation is the same as what has been known for the nucleon case ($N_f=2$) 
\cite{Hashimoto:2010je}.

In the next section, we review briefly the recent lattice calculations of the inter-baryon potential for three-flavor QCD,
and discuss a comparison with our holographic result.

\section{A comparison with lattice QCD and OPE}

In the previous section, we have calculated a short-distance potential between two baryons in multi-flavor holographic QCD.
We have found a universal repulsive potential for generic baryon states.
In this section, we shall compare our results with ones obtained by
a completely different technique: the lattice QCD.

First, we shall review the results in lattice QCD.
Potentials between two octet baryons have been investigated in lattice QCD in the flavor SU(3) symmetric limit \cite{Inoue:2010hs,Inoue:2011ai}, where all quark mass in the 3 flavor QCD are artificially taken to be equal, $m_u=m_d=m_s$, with the lattice spacing $a\simeq 0.12$ fm and the spatial extension $L\simeq 2 - 4$ fm.  
Simulations employ six different values of quark mass, which correspond to 
the pseudo-scalar  meson mass $m_{\rm PS}\simeq $ 470, 670, 840, 1020, 1170 MeV, 
where the relation that $m_{\rm PS}^2 = A m_q $ holds for a small quark mass $m_q$  with a common coefficient $A$. 
There are 6 independent potentials between two octet baryons, which correspond to
 irreducible representations of the flavor SU(3) group  as
\begin{equation}
{\bf 8}\otimes {\bf 8} = \underbrace{{\bf 27}\oplus {\bf 8}_s \oplus {\bf 1}}_{\rm symmetric} \oplus \overbrace{{\bf 10}^* \oplus {\bf 10} \oplus {\bf 8}_a}^{\rm anti-symmetric} ,
\end{equation}
where the first 3 representations are symmetric under the exchange of two octet baryons, while
the last 3 are anti-symmetric. To satisfy a condition that a total wave function is odd under exchange of two octet baryons, the first 3 states have spin zero ($S=0$, odd under the exchange) while the last 3 states must have spin one ($S=1$, even under the exchange) if the orbital angular momentum between two baryons  is zero ($L=0$).
 
\begin{figure}[htb]
\centering
 \includegraphics[width=0.49\textwidth]{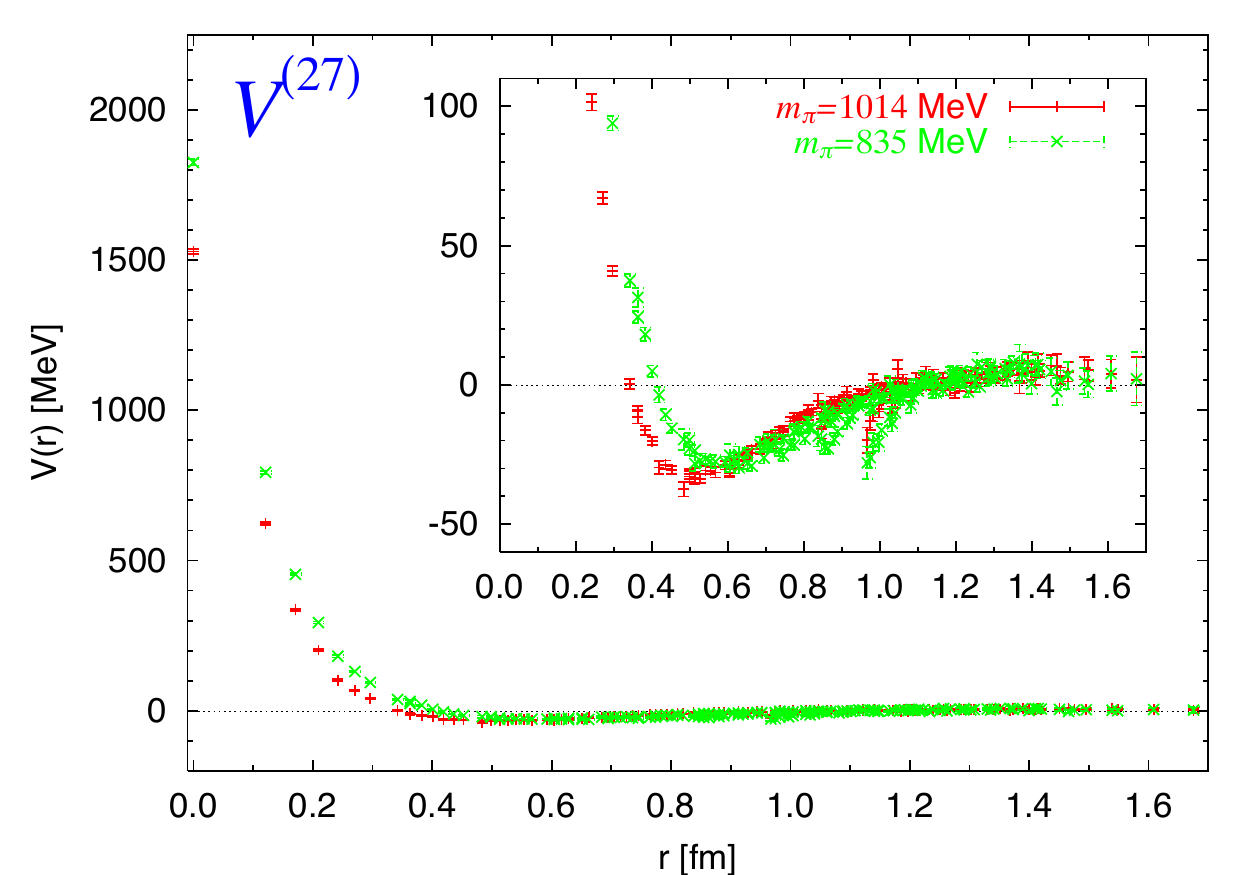}\hfill
 \includegraphics[width=0.49\textwidth]{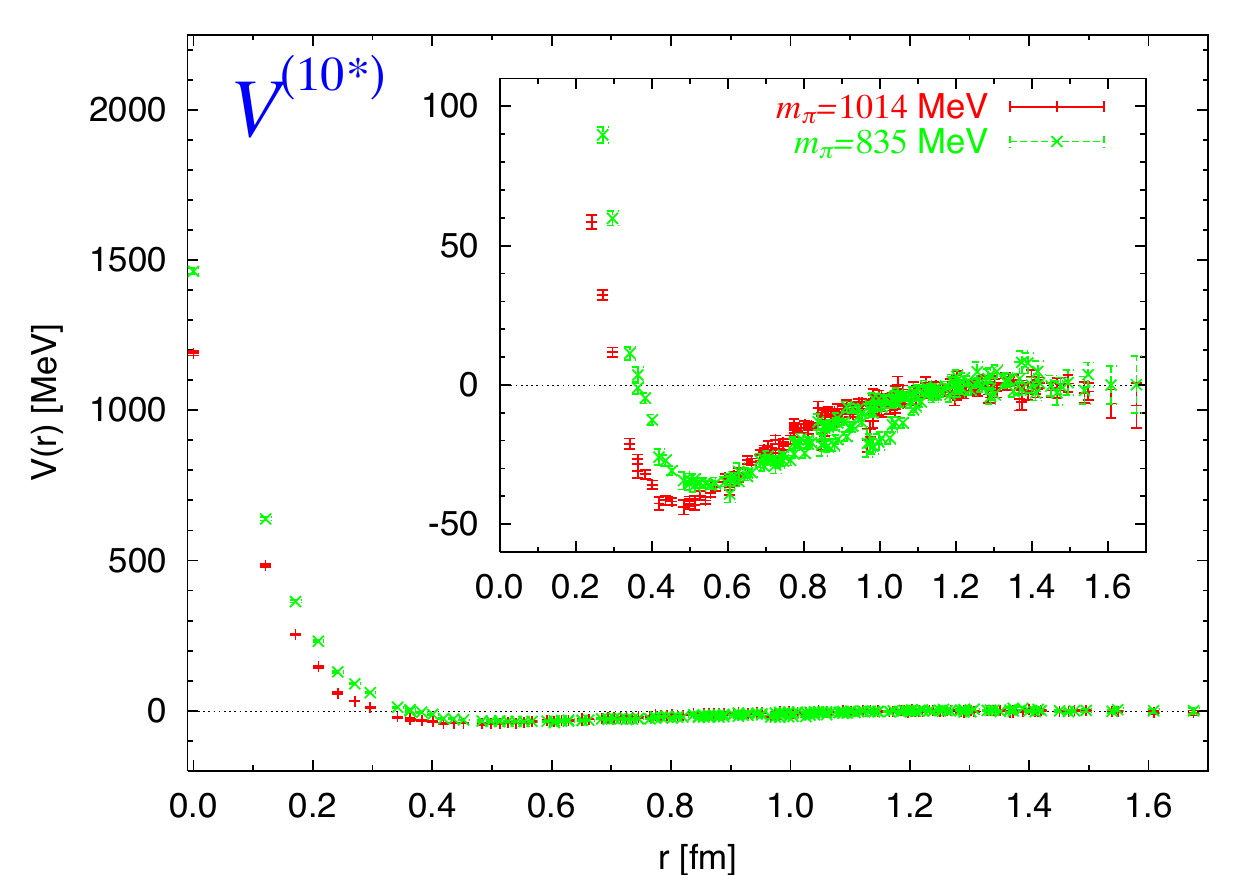}
 \includegraphics[width=0.49\textwidth]{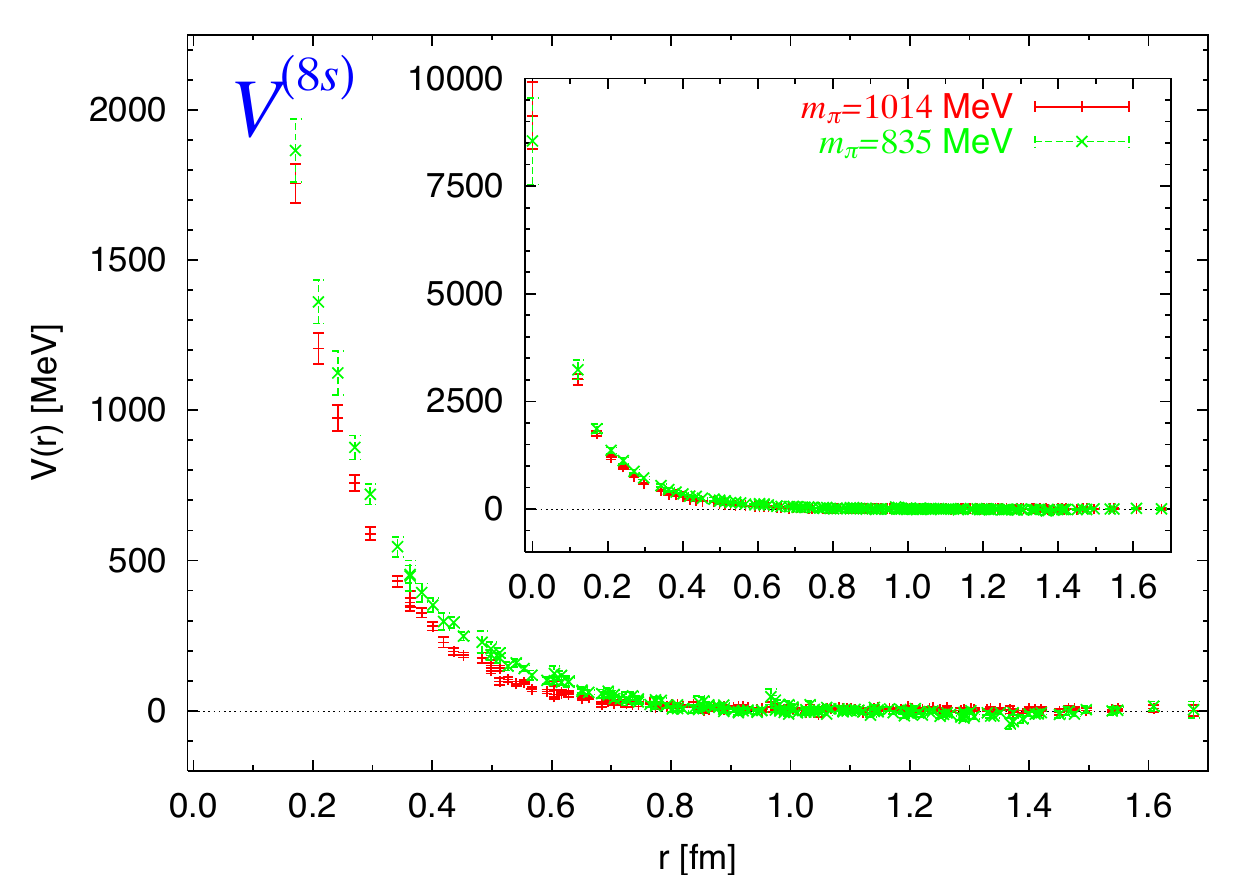}\hfill
 \includegraphics[width=0.49\textwidth]{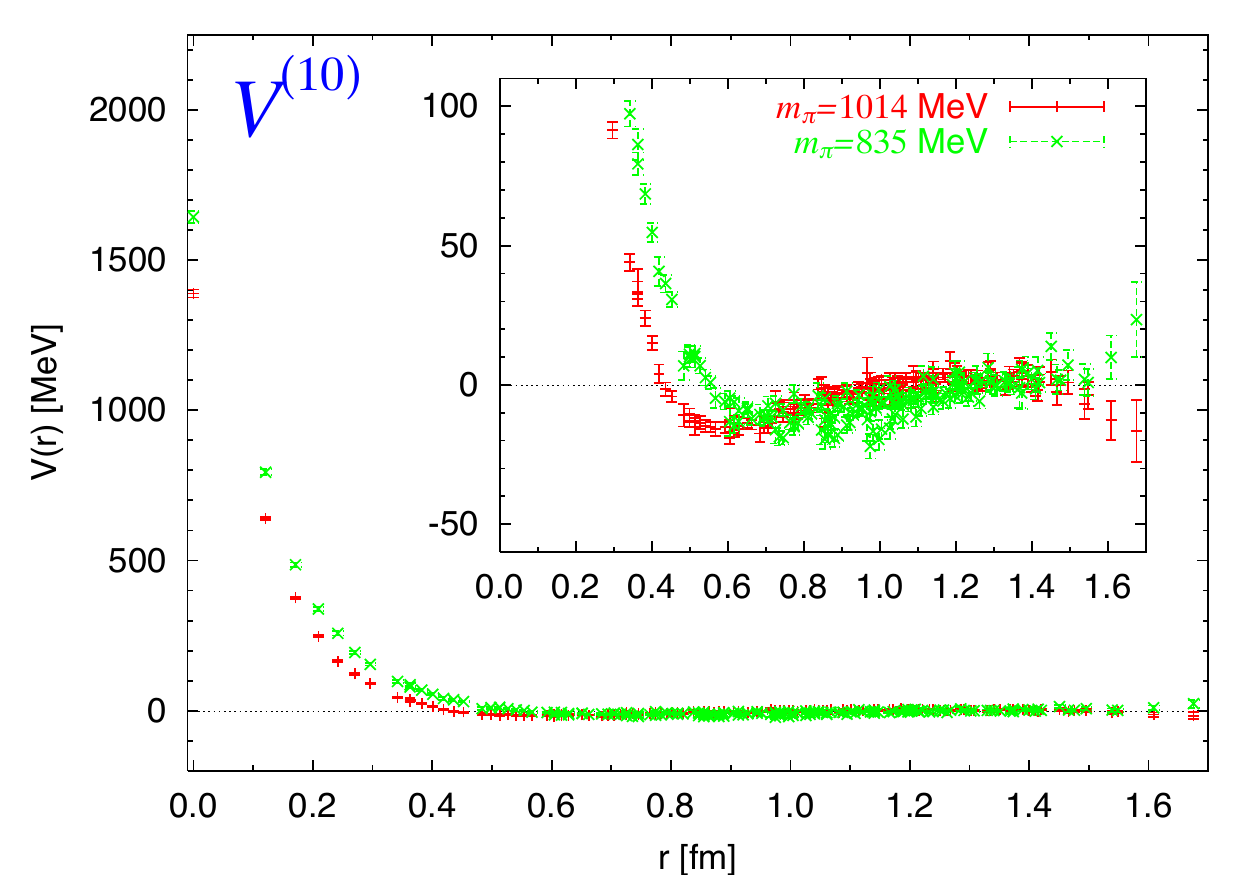}
 \includegraphics[width=0.49\textwidth]{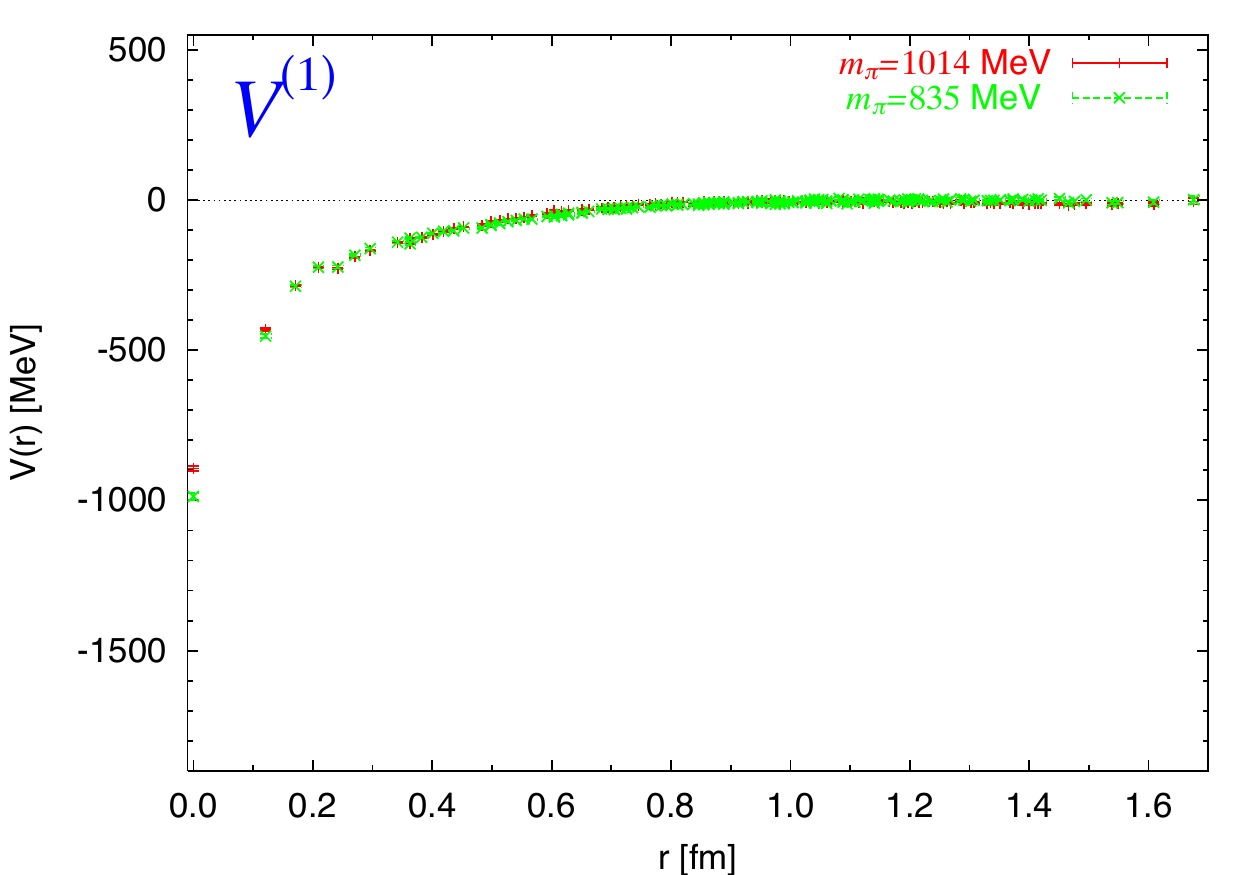}\hfill
 \includegraphics[width=0.49\textwidth]{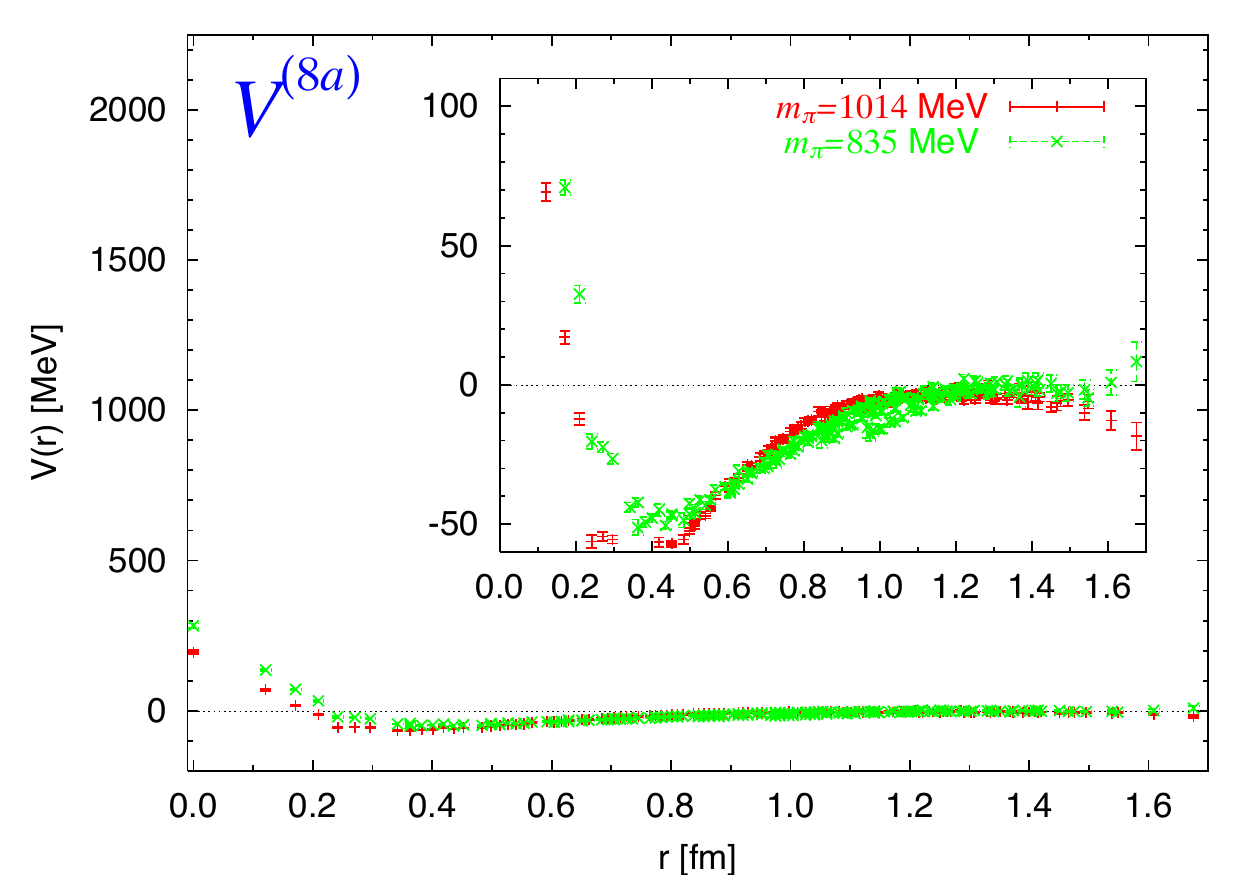}
\caption{The 6 independent potentials in the flavor SU(3) limit obtained in lattice QCD at $m_{\rm PS}=1014$ MeV (red) and 835 MeV (green) \cite{Inoue:2010hs}. }
\label{fig:BBpotential}
\end{figure}

A typical example of corresponding potentials is shown in Fig.~\ref{fig:BBpotential},
taken from Ref.~\cite{Inoue:2010hs}, where the central potentials (left three) and the effective central potential
(right three), where an effect of the tensor potential is included, 
 are plotted. 

\begin{figure}[htb]
\centering
 \includegraphics[width=0.49\textwidth]{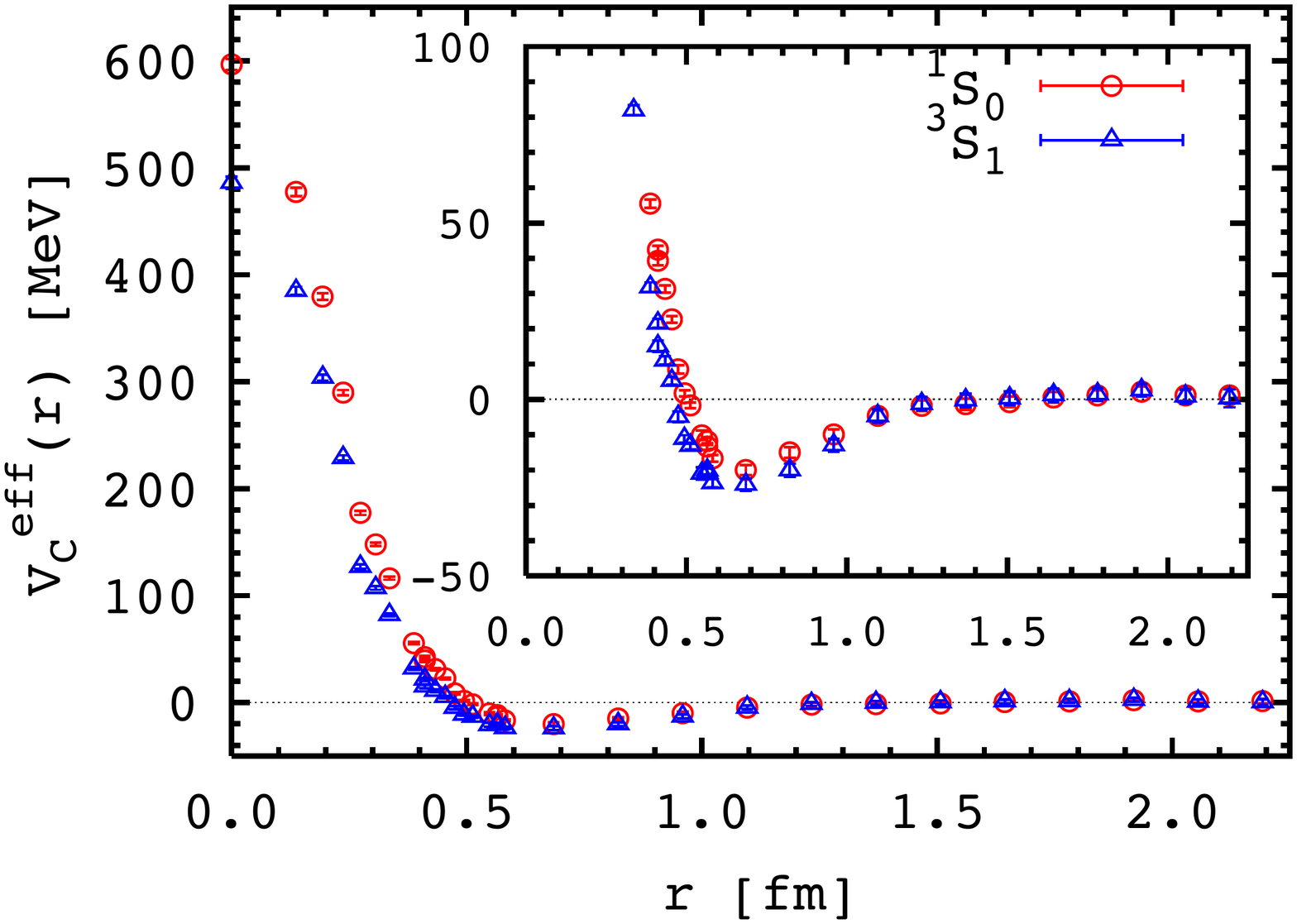}
\caption{NN potentials in quenched QCD at $m_{\pi}\simeq 730$ MeV \cite{Aoki:2009ji}. 
 The spin singlet sector ($^1S_0$)
belongs to the ${\bf 27}$ representation while the triplet to the ${\bf 10}^*$ in the flavor SU(3). }
\label{fig:NNpotential}
\end{figure}

As can be seen from Fig.~\ref{fig:BBpotential}, inter-baryon potentials strongly depend on the representations.  In top panels, $V^{({\bf 27})}$ and $V^{({\bf 10}^*)}$, which correspond to isospin-triplet and isospin-singlet nucleon-nucleon (NN) potentials in the $N_f=2$ case\footnote{This is because 
the Young tableau of both $27$ and $10^*$ do not have three rows.} \cite{Ishii:2006ec,Aoki:2009ji}, respectively, have a repulsive core at short distance and  an attractive pocket at medium distance. These features qualitatively agree with those of  the NN potentials in quenched QCD,  shown in Fig.~\ref{fig:NNpotential}.
For $L=0$,  $V^{({\bf 27})}$ is isospin-triplet ($I=1$) at $N_f = 2$ and spin-singlet ($S=0$) while
 $V^{({\bf 10}^*)}$ is isospin-singlet ($I=0$) at $N_f=2$ and spin-triplet ($S=1$). 
 Therefore, the flavor singlet potential at $N_f=2$ can not have spin-zero for $L=0$.
 
On the other hand if the strange quark is introduced in the flavor representation, we 
 have more varieties of potentials: $V^{({\bf 10})}$ has a stronger repulsive core and a weaker attractive pocket than $V^{({\bf 27})}, V^{({\bf 10}^*)}$, and $V^{({\bf 8}_s)}$ has only a repulsion with the strongest repulsive core among all, while $V^{({\bf 8}_a)}$  has a strongest attractive pocket with the weakest repulsive core. In contrast to these five cases, the singlet potential, $V^{({\bf 1})}$ shows attraction at all distances without repulsive core, which produces one bound state, the H-dibaryon, in this channel \cite{Inoue:2010es, Beane:2010hg}. Note that the flavor singlet potential has spin-zero for $L=0$ in this case, contrary to the $N_f=2$ case.

Increasing the number of flavor from 2 to 3, we observe that repulsive core becomes weaker in some channel (${\bf 8}_a$) and it even disappears in the singlet (${\bf 1}$), as seen in Tab.~\ref{tab:BBpotentials}, where we summarize features of inter-baryonic potential in the flavor SU(3) limit.

Now let us discuss a comparison between our holographic QCD results and the lattice QCD results.
In the previous section, we have found a universal repulsive core for multi-flavor inter-baryon potential.
On the other hand, in the three-flavor lattice QCD, in most of the channels there appears repulsive cores.
Therefore we conclude that our holographic results are consistent with the lattice QCD results, generically. 

Only one exception is the existence of an attractive channel. 
In the lattice QCD result, the flavor-singlet combination of the baryons in the ${\bf 8}$ representation 
for $N_f=3$ is found to have a vanishing repulsive core. In the holographic side, as we work with
the large $N_c$, it is not clear how the lattice QCD with $N_f=N_c=3$ can be mapped to the 
holographic QCD. However, in the previous section, we have seen that a classical 
inter-baryon potential can vanish. So the disappearance of the
repulsive core in the lattice QCD is not a contradiction with the holographic QCD. We leave
a more detailed comparison to a future work.

In table~\ref{tab:BBpotentials}, we summarize qualitative features of baryon-baryon potentials, together with the prediction from the operator product expansion (OPE) in perturbative QCD for their short distance behaviors \cite{Aoki:2010kx, Aoki:2009pi,Aoki:2010uz}. Although the OPE analysis is consistent with  
the attractive core for the singlet potential in the lattice QCD, 
it  disagrees with  
the strong repulsion of the ${\bf 8}_s$ potential  in the lattice QCD.\footnote{
This disagreement  
may be related to the fact that non-relativistic 6 quark operators are absent in this channel.}
Obviously it is 
desirable to investigate the short distance behaviors of the inter-baryon potential by various methods, including holographic  QCD, in more details.

\begin{table}[htdp]
\begin{center}
\begin{tabular}{|c||c|c|c|c|c|c|}
\hline
representation & ${\bf 27}$ & ${\bf 8}_s$ & ${\bf 1}$ & ${\bf 10}^*$ & ${\bf 10}$ & ${\bf 8}_a$ \\
\hline
repulsion & yes & strongest & no & yes & strong & weak \\
attraction  & yes & no & strongest & yes & weak & strong \\
comment & $NN(I=1)$& & H-dibaryon & $NN(I=0)$ &  &  \\
\hline
OPE & rpl. & att. & atr. & rpl. & rpl. & atr. \\ 
\hline
\end{tabular}
\caption{Overall feature of inter-baryon potential in each representation. The last line shows the short distance behavior of the potential from OPE, where rpl.=repulsive and atr.=attractive. }
\end{center}
\label{tab:BBpotentials}
\end{table}%

\acknowledgments
We would like to thank T.~Hatsuda, M.~Hidaka, T.~Morita, K.~Yazaki, and P.~Yi. 
for valuable discussions.
S.~A.~is supported in part by  Grant-in-Aid for Scientific Research on 
Innovative Areas (No. 2004: 20105001,20105003) and by SPIRE (Strategic Program for Innovative REsearch).
This research was partially supported by 
KAKENHI Grant-in-Aid 23654096, 23105716, 22340069, 21105514.
K. H.~is partly supported by
the Japan Ministry of Education, Culture, Sports, Science and
Technology. N.I. would like to thank mathematical physics laboratory at RIKEN for very kind hospitality where 
part of this work is done.


\end{document}